\pdfoutput=1
\documentclass[aps,prd,amsmath,floats,floatfix, twocolumn,
superscriptaddress,nofootinbib,showpacs,longbibliography]{revtex4-1}
\usepackage[T1]{fontenc}
\usepackage[utf8]{inputenc}
\usepackage{lmodern}

\usepackage{verbatim}

\usepackage[dvipsnames, usenames]{xcolor}
\definecolor{linkcolor}{rgb}{0.0,0.3,0.5}
\usepackage[hypertexnames=false, unicode, colorlinks=true, linkcolor=linkcolor,
citecolor=linkcolor, filecolor=linkcolor,urlcolor=linkcolor,
pdfusetitle]{hyperref}

\usepackage[all]{hypcap}
\usepackage{graphicx}
\usepackage{xspace}
\usepackage{amssymb}
\usepackage[normalem]{ulem}
\usepackage{bm}

\usepackage{microtype}

\usepackage[english]{babel}
\usepackage{blindtext}

\newcommand\prlsec[1]{\vspace{2mm}\noindent \textbf{\emph{#1}}---}

\graphicspath{%
  {figs/}%
}

\DeclareMathAlphabet{\mathpzc}{OT1}{pzc}{m}{it}

\newcommand{\roughly}{\mathchar"5218\relax\,}

\newcommand{\Li}{\mathcal{L}}

\newcommand{\bchi}{\bm{\chi}}
\newcommand{\blambda}{\bm{\lambda}}
\newcommand{\bLambda}{\bm{\Lambda}}

\newcommand{\bv}{\bm{v}}

\newcommand{\trefmHundredM}{t_{\mathrm{ref}}/M\!=\!-100}
\newcommand{\frefTwentyHz}{f_{\mathrm{ref}}\!=\!20 \, \mathrm{Hz}}
\newcommand{\vesc}{v^{\mathrm{max}}_{\mathrm{esc}}}
\newcommand{\vLower}{$v_f \gtrsim 698$ km/s\xspace}
\newcommand{\vRange}{$v_f \sim 1542^{+747}_{-1098}$ km/s\xspace}
\newcommand{\vRangeBare}{$\roughly 1542^{+747}_{-1098}$ km/s\xspace}
\newcommand{\KLDiv}{4.3\xspace}

\newcommand{\NRSur}{\texttt{NRSur7dq4}\xspace}
\newcommand{\NRSurRemnant}{\texttt{NRSur7dq4Remnant}\xspace}
\newcommand{\EOB}{\texttt{SEOBNRv4PHM}\xspace}
\newcommand{\Phenom}{\texttt{IMRPhenomXPHM}\xspace}

\newcommand{\Cornerfignum}{1\xspace}

\newcommand{\AEI}{\affiliation{Max Planck Institute for Gravitational Physics
    (Albert Einstein Institute), Am M\"uhlenberg 1, Potsdam 14476, Germany}} %

\newcommand\MIT{\affiliation{LIGO Laboratory, Massachusetts Institute of
Technology, Cambridge, Massachusetts 02139, USA}}
\newcommand{\MKI}{\affiliation{Department of Physics and Kavli Institute for Astrophysics and Space Research, Massachusetts Institute of Technology, 77 Massachusetts Ave, Cambridge, MA 02139, USA}}
\newcommand{\CCA}{\affiliation{Center for Computational Astrophysics, Flatiron Institute, New York NY 10010, USA}}
\newcommand{\StonyBrook}{\affiliation{Department of Physics and Astronomy, Stony Brook University, Stony Brook NY 11794, USA}}
\newcommand{\UMassDMath}{\affiliation{Department of Mathematics,
		University of Massachusetts, Dartmouth, MA 02747, USA}}
\newcommand{\CSCVRUMass}{\affiliation{Center for Scientific Computing and Data Science Research, University of Massachusetts, Dartmouth, MA 02747, USA}}

\newcommand{\TITLE}{Evidence of large recoil velocity from a black hole merger
signal}

\begin{document}

\title{\TITLE}

\author{Vijay Varma}
\email{vijay.varma@aei.mpg.de}
\thanks{Marie Curie Fellow}
\AEI

\author{Sylvia Biscoveanu}
\MIT
\MKI

\author{Tousif Islam}
\UMassDMath
\CSCVRUMass

\author{Feroz H. Shaik}
\UMassDMath
\CSCVRUMass

\author{Carl-Johan Haster}
\MIT
\MKI

\author{Maximiliano Isi}
\CCA

\author{Will M. Farr}
\StonyBrook
\CCA

\author{Scott E. Field}
\UMassDMath
\CSCVRUMass

\author{Salvatore Vitale}
\MIT
\MKI

\hypersetup{pdfauthor={Varma el al.}}

\date{\today}

\begin{abstract}
The final black hole left behind after a binary black hole merger can attain a
recoil velocity, or a ``kick'', reaching values up to 5000 km/s. This
phenomenon has important implications for gravitational wave astronomy, black
hole formation scenarios, testing general relativity, and galaxy evolution. We
consider the gravitational wave signal from the binary black hole merger
GW200129\!{\tiny {\bf \_}}065458 (henceforth referred to as GW200129), which
has been shown to exhibit strong evidence of orbital precession. Using
numerical relativity surrogate models, we constrain the kick velocity of
GW200129 to $v_f \sim 1542^{+747}_{-1098}$ km/s or $v_f \gtrsim 698$ km/s
(one-sided limit), at 90\% credibility. This marks the first identification of
a large kick velocity for an individual gravitational wave event. Given the
kick velocity of GW200129, we estimate that there is a less than $0.48\%$
($7.7\%$) probability that the remnant black hole after the merger would be
retained by globular (nuclear star) clusters.  Finally, we show that kick
effects are not expected to cause biases in ringdown tests of general
relativity for this event, although this may change in the future with improved
detectors.
\end{abstract}

\maketitle

\prlsec{Introduction.}
When two black holes (BHs) orbit each other, they emit gravitational waves
(GWs) which carry away energy and angular momentum. This causes the orbit to
shrink in a runaway process that culminates in the merger of the BHs into a
single remnant BH. At the same time, the GWs can also carry away linear
momentum from the binary, shifting its center of mass in the opposite
direction~\cite{Fitchett:1983MNRAS}. Most of the linear momentum is lost near
the merger~\cite{Gonzalez:2006md}, resulting in a recoil or ``kick'' velocity
imparted to the remnant BH.

Kicks are particularly striking for precessing binaries, in which the component
BH spins are tilted with respect to the orbital angular momentum. For these
systems, the spins interact with the orbital angular momentum as well as with
each other, causing the orbital plane to precess~\cite{Apostolatos:1994pre}.
Numerical relativity (NR) simulations revealed that the kick velocities for
precessing binaries can reach values up to $\sim5000$
km/s~\cite{Campanelli:2007cga, Gonzalez:2007hi, Lousto:2011kp}, large enough to
be ejected from any host galaxy~\cite{Merritt:2004xa}.

Kicks have important implications for BH astrophysics. Following a supermassive
BH merger, the remnant BH can be displaced from the galactic center or ejected
entirely~\cite{Merritt:2004xa}, impacting the galaxy's
evolution~\cite{Komossa:2008as}, fraction of galaxies with central supermassive
BHs~\cite{Volonteri:2010mbh}, and event rates~\cite{sesana:2007zk} for the
future LISA mission~\cite{AmaroSeoane:2017las}. For stellar-mass BHs like those
observed by LIGO~\cite{TheLIGOScientific:2014jea} and
Virgo~\cite{TheVirgo:2014hva}, kicks can limit the formation of heavy BHs. BH
masses greater than ${\sim}65 M_{\odot}$ are disfavored by supernova
simulations~\cite{Woosley:2016hmi, Marchant:2018kun}, but have been seen in GW
events~\cite{Abbott:2020tfl, Abbott:2020niy, LIGOScientific:2021djp}. This
could be explained by second-generation mergers~\cite{Gerosa:2021mno}, in which
one of the component BHs is itself a remnant from a previous merger, and is
thus more massive than the original stellar-mass progenitors. However, if the
kick from the first merger is large enough, the remnant BH would get ejected
from its host galaxy and would not participate in another merger.

Unfortunately, observational evidence of large kicks has been elusive. While
various candidates from electromagnetic observations have been identified,
their nature is debated~\cite{Bogdanovic:2021aav}. Similarly, observing kicks
using GW signals has been challenging~\cite{Varma:2020nbm, Abbott:2020mjq,
Gerosa:2016vip, CalderonBustillo:2018zuq, Lousto:2019lyf, Healy:2020jjs,
Mahapatra:2021hme}.  For example,
\citeauthor{Varma:2020nbm}~\cite{Varma:2020nbm} used accurate models based on
NR simulations to show that kicks from precessing binaries can be reliably
inferred with LIGO-Virgo operating at their design sensitivity.  However, the
GW events analyzed in Ref.~\cite{Varma:2020nbm}, which only included signals in
the first two LIGO-Virgo observing runs~\cite{LIGOScientific:2018mvr}, were not
loud enough to constrain the kick.

Since then, the LIGO-Virgo detectors have been further upgraded, and the GW data
from the third observing run were released in two stages,
O3a~\cite{Abbott:2020niy} and O3b~\cite{LIGOScientific:2021djp}. Notably, O3a
provided the first evidence for precession in the ensemble population of merging
binaries~\cite{Abbott:2020gyp}, even though none of the individual GW events
unambiguously exhibited precession~\cite{Abbott:2020niy}. Finally, in O3b, the
binary BH merger GW200129 was identified as the first individual GW event
showing strong evidence of precession~\cite{LIGOScientific:2021djp,
Hannam:2021pit}. Similarly, support for large kicks was identified in the
ensemble population using the O3a data~\cite{Varma:2021xbh, Doctor:2021qfn},
even though the individual events were not loud enough for an unambiguous kick
inference~\cite{Varma:2020nbm, Islam:2022_inprep} (with the exception of
GW190814~\cite{Abbott:2020khf}, which was found to have a \emph{small} kick of
$\sim 74^{+10}_{-7}$ km/s at 90\% credibility~\cite{Mahapatra:2021hme}).

In this \emph{Letter}, we use the method developed in Ref.~\cite{Varma:2020nbm}
to show that GW200129 has a large kick velocity (\vRangeBare at 90\%
credibility).  As an application of the kick constraint, we compute the
retention probability for the remnant BH of GW200129 in various host
environments, and discuss the implications for the formation of heavy
stellar-mass BHs. Finally, we show that Doppler effects due to the kick on the
remnant mass measurement are small for this event, and should not impact
ringdown tests of general relativity (GR).

\prlsec{Methods.}
We follow the procedure outlined in Ref.~\cite{Varma:2020nbm} to infer the kick
from a GW signal. We begin by measuring the binary source parameters following
Bayes' theorem~\cite{Thrane:2019pe}:
\begin{gather}
    p(\blambda|d) \propto \Li(d|\blambda) \, \pi(\blambda),
\label{eq:Bayes}
\end{gather}
where $p(\blambda|d)$ is the \emph{posterior} probability distribution of the
binary parameters $\blambda$ given the observed data $d$, $\Li(d|\blambda)$ is
the \emph{likelihood} of the data given $\blambda$, and $\pi(\blambda)$ is the
\emph{prior} probability distribution for $\blambda$. Under the assumption of
Gaussian detector noise, the likelihood $\Li(d|\blambda)$ can be evaluated for
any $\blambda$ using a gravitational waveform model and the observed data
stream $d$~\cite{Thrane:2019pe}. A stochastic sampling algorithm is then used
to draw \emph{posterior samples} for $\blambda$ from $p(\blambda|d)$.  We use
the \texttt{Parallel Bilby}~\cite{Smith:2019ucc} parameter estimation package
with the \texttt{dynesty}~\cite{Speagle:2019dynesty} sampler.

For quasicircular binary BHs, the full set of parameters $\blambda$ is 15
dimensional~\cite{LIGOScientific:2021djp}. This includes the 8D intrinsic
parameters: the component masses ($m_1$ and $m_2$) and spins ($\bchi_{1}$ and
$\bchi_{2}$, each of which is a 3D vector), as well as the 7D extrinsic
parameters: the distance, right ascension, declination, time of arrival,
coalescence phase, binary inclination, and polarization angle. Here, index 1
(2) corresponds to the heavier (lighter) BH, $\bchi_{1,2}$ are dimensionless
spins with magnitudes $\chi_{1,2} \leq 1$, and masses refer to the detector
frame redshifted masses. We also define the mass ratio $q=m_1/m_2 \geq 1$,
total mass $M=m_1+m_2$, and use geometric units with $G=c=1$.

We employ the NR surrogate models \NRSur~\cite{Varma:2019csw} and
\NRSurRemnant~\cite{Varma:2019csw, Varma:2018aht} to infer the kick.
Constructed by effectively interpolating between $\sim1500$ precessing NR
simulations, \NRSur predicts the gravitational waveform, while \NRSurRemnant
predicts the mass $m_f$, spin $\bchi_f$, and kick velocity $\bv_f$ of the
remnant BH. We first obtain posterior samples for all 15 binary parameters
using \NRSur. The spins are measured in a source frame defined at a given
reference point (see below): the $z$-axis lies along the instantaneous orbital
angular momentum, the $x$-axis points along the line of separation from the
lighter to the heavier BH, and the $y$-axis completes the right-handed triad.
The remnant properties, which are also defined in the same source frame, depend
only on the intrinsic parameters $\bLambda=\{m_1, m_2, \bchi_1, \bchi_2\}$.
Therefore, the posteriors for the remnant properties are obtained by evaluating
\NRSurRemnant on the $\bLambda$ posterior samples; put simply, \NRSurRemnant is
a function of $\bLambda$ that yields $m_f$, $\bchi_f$ and $\bv_f$. We also
compute the \emph{effective prior} distribution for $\bv_f$, by evaluating
\NRSurRemnant on $\bLambda$ samples drawn from the prior $\pi(\bLambda)$. The
difference between the kick posterior and prior can be used to gauge how
informative the data are about the kick~\cite{Varma:2020nbm}.

Traditional modeling methods assume a phenomenological ansatz for the
waveform~\cite{Pratten:2020ceb, Ossokine:2020kjp} or remnant
properties~\cite{Hofmann:2016yih, Barausse:2012qz, Jimenez-Forteza:2016oae,
Lousto:2012su}, and calibrate remaining free parameters to NR simulations. NR
surrogate methods~\cite{Varma:2019csw, Varma:2018aht, Varma:2018mmi,
Blackman:2017pcm}, on the other hand, take a data-driven approach and the
models are trained directly against precessing NR simulations. In this
approach, one first constructs a suitable numerical basis using a subset of the
NR waveforms, and then builds fits across parameter space for the basis
coefficients; we refer the reader to Ref.~\cite{Varma:2019csw} for more
details. NR surrogate models do not need to introduce additional assumptions
about the underlying phenomenology which would necessarily introduce some
systematic error. Through cross-validation studies, it has been shown that both
\NRSur and \NRSurRemnant achieve accuracies comparable to the simulations
themselves~\cite{Varma:2019csw}, and as a result, are the most accurate models
currently available for precessing systems, within their parameter space of
validity: both models are trained on simulations with $q\leq4$ and $\chi_{1,2}
\leq 0.8$, but can be extrapolated to $q\leq6$ and
$\chi_{1,2}\leq1$~\cite{Varma:2019csw}. As GW200129 shows significant support
for large spins, we conduct some tests of the surrogate models in this regime
in the Supplement~\cite{kickpapersupplement}.

For the prior in Eq.~(\ref{eq:Bayes}), we follow
Ref.~\cite{LIGOScientific:2021djp} and adopt a uniform prior for spin
magnitudes (with $0 \leq \chi_{1},\chi_{2}\leq 0.99$) and redshifted component
masses, an isotropic prior for spin orientations, sky location and binary
orientation, and a distance prior
(\texttt{UniformSourceFrame}~\cite{Romero-Shaw:2020owr}) that assumes uniform
source distribution in comoving volume and time. In addition, we place the
following constraints: $q \leq 6$ and $60\leq M \leq 400$. These constraints
are motivated by the regime of validity of \NRSur~\cite{Varma:2019csw}, and are
broad enough to safely encompass the posterior spread of
GW200129~\cite{LIGOScientific:2021djp}.

\begin{figure}[thb]
\includegraphics[width=0.49\textwidth]{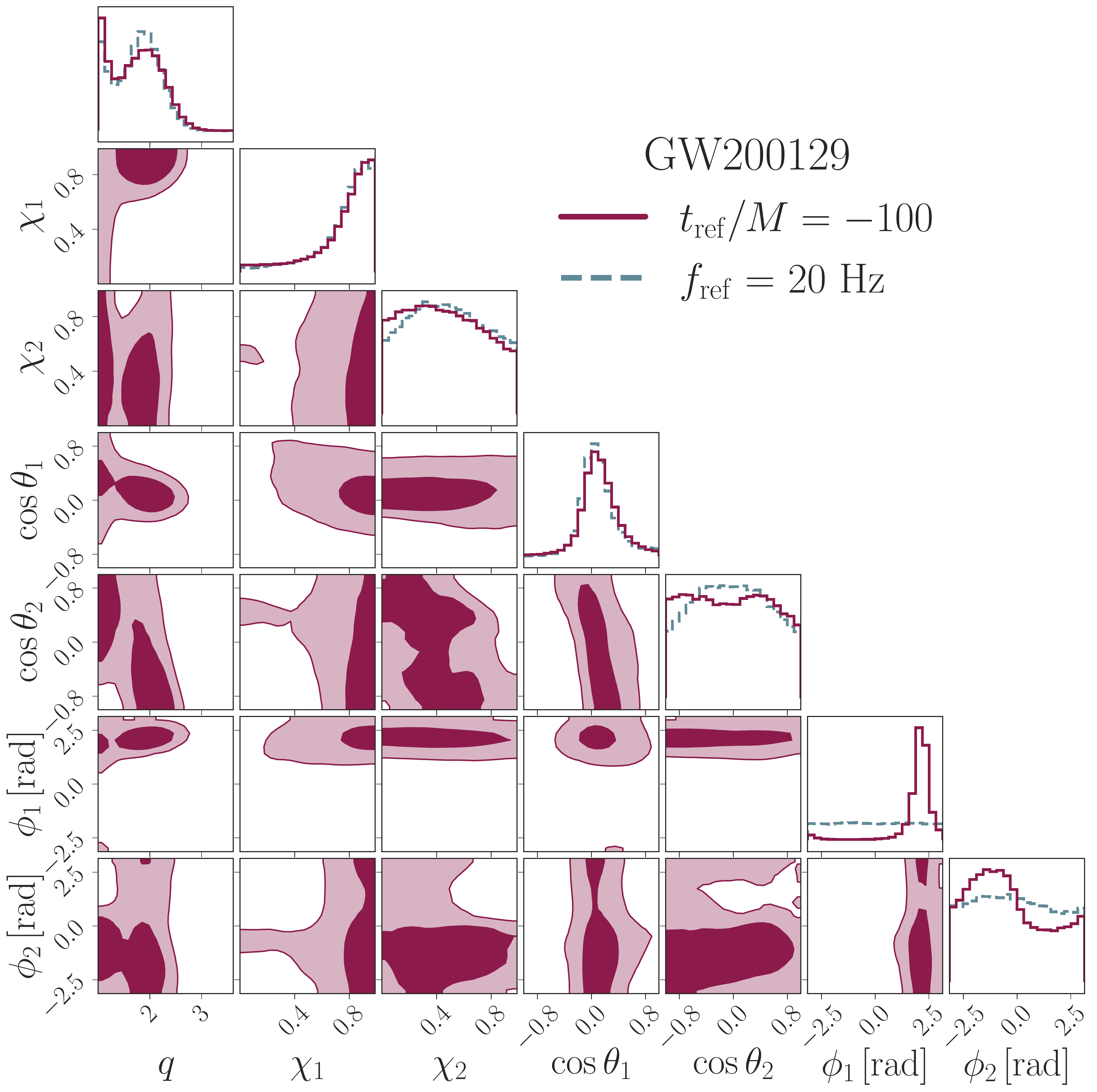}
\caption{
Constraints on the mass ratio and spins for GW200129, at reference time
$\trefmHundredM$. The dark (light) regions represent the 50\% (90\%) credible
bounds on joint 2D posteriors, while the diagonal plots show 1D marginalized
posteriors. There is a preference for large $\chi_1$ and $\cos\theta_1 \sim 0$,
meaning there is substantial spin in the orbital plane, which leads to
precession. For comparison, we also show the 1D marginalized posteriors at
$\frefTwentyHz$. The azimuthal spin angles (especially
$\phi_1$) are much better constrained at $\trefmHundredM$; this is critical for
constraining the kick.
}
\label{fig:corner_GW200129}
\end{figure}

Because the spin directions are not constant for precessing binaries, spin
measurements are inherently tied to a specific moment in the binary's
evolution. The standard approach is to measure the spins at the point where
the frequency of the GW signal at the detector reaches a prespecified reference
value, typically $\frefTwentyHz$~\cite{Abbott:2020niy}. This is mainly
motivated by the fact that the sensitivity band of current detectors begins
near this value~\cite{TheLIGOScientific:2014jea, TheVirgo:2014hva}.  However,
Ref.~\cite{Varma:2021csh} recently showed that constraints on orbital-plane
spin directions can be greatly improved by measuring the spins near the merger,
in particular, at a fixed \emph{dimensionless} reference time $\trefmHundredM$
before the peak of the GW amplitude.  This improvement can be attributed to the
waveform being more sensitive to variations in the orbital-plane spin
directions near the merger~\cite{Varma:2021csh} ($\trefmHundredM$ typically
falls within $\sim 2-4$ GW cycles before the peak amplitude, independent of the
binary parameters~\cite{Varma:2021csh}.).

We will adopt the $\trefmHundredM$ reference point for the main results in this
paper, but will show a comparison against $\frefTwentyHz$ for completeness.  As
we will discuss below, the choice of reference point has a negligible impact on
the kick inference itself, but comparing the spin posteriors at the two
reference points helps illustrate why a kick constraint is possible in the
first place.  As a bonus, spin measurements at $\trefmHundredM$ are convenient
for inferring the kick as the \NRSurRemnant model is also trained at this
reference time~\cite{Varma:2019csw}; this choice was found to lead to a more
accurate remnant BH model in Refs.~\cite{Varma:2019csw, Varma:2018aht}.

\begin{figure*}[thb]
\includegraphics[width=0.49\textwidth]{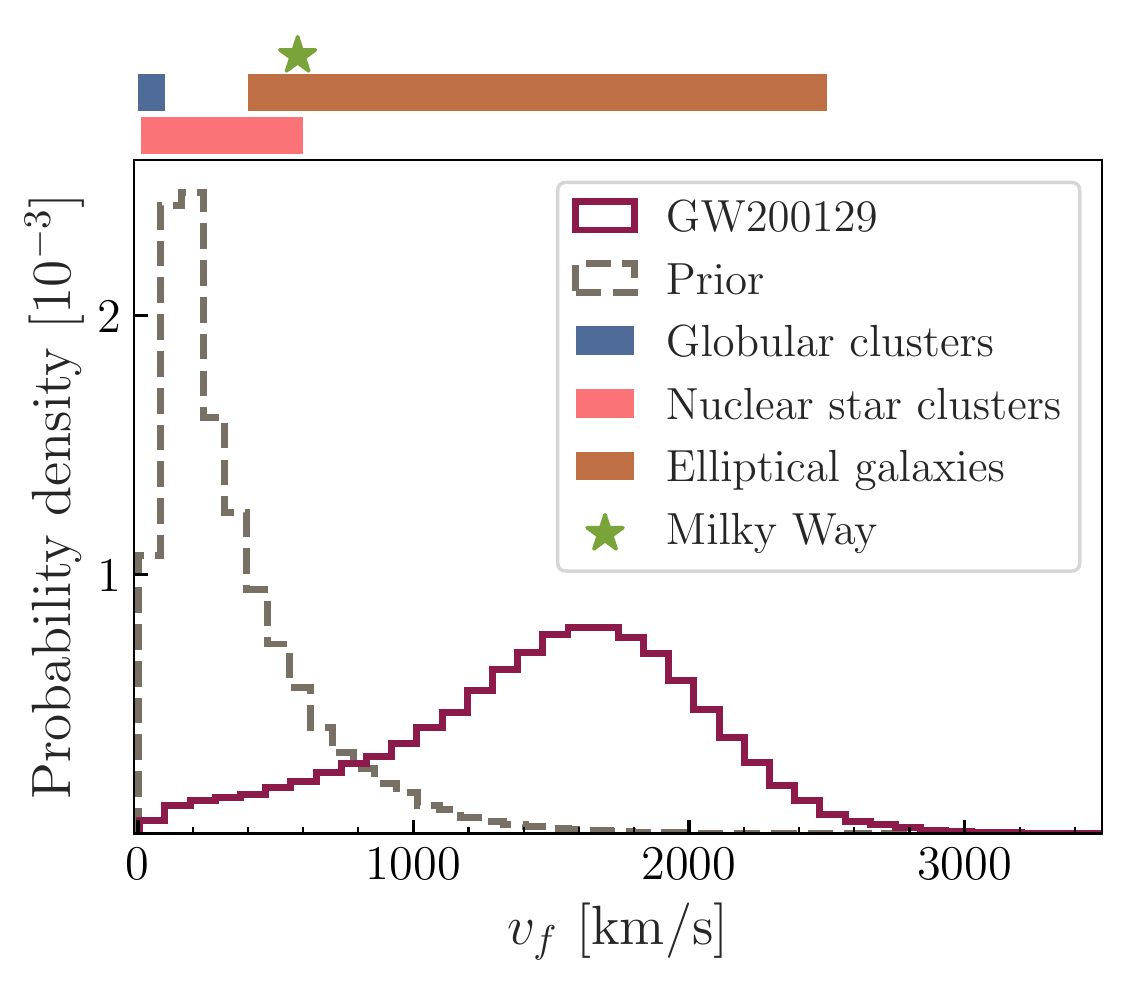} ~\,
\includegraphics[width=0.49\textwidth]{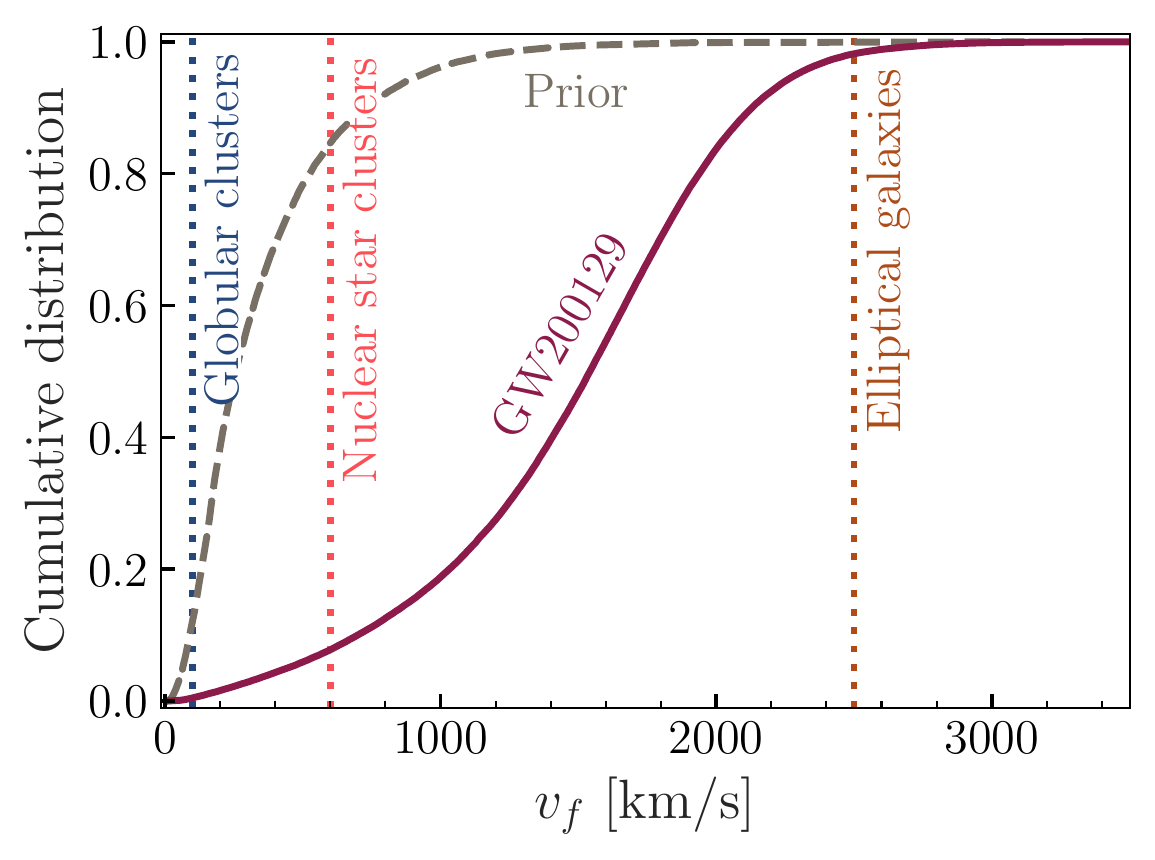}
\caption{
\emph{Left:} Kick magnitude constraints for GW200129. We show the posterior and
the effective prior, along with known ranges for the escape velocities for
various types of host environments for comparison. There is a clear preference
for large kicks in the posterior, with \vLower at 90\% credibility.
\emph{Right:} Cumulative distribution functions (CDFs) for the kick posterior
and prior. The upper bounds of the escape velocity ranges from the left panel
are shown as vertical dotted lines. The upper limit for retention probability
of the merger remnant is given by the intersection of these lines with the
posterior CDF.
}
\label{fig:kick}
\end{figure*}

\prlsec{GW200129 spin measurements.}
GW200129 is the first GW event showing strong signs of
precession~\cite{LIGOScientific:2021djp, Hannam:2021pit}.
Figure~\ref{fig:corner_GW200129} shows the posterior distribution for the mass
ratio and spin parameters obtained using the \NRSur model at reference points
$\trefmHundredM$ and $\frefTwentyHz$; our constraints at $\frefTwentyHz$ are
consistent with those of Ref.~\cite{Hannam:2021pit}.  The spin vectors
$\bchi_{1,2}$ are decomposed into magnitudes $\chi_{1,2}$, tilts angles
$\theta_{1,2}$ with respect to the $z$-axis, and azimuthal angles $\phi_{1,2}$
with respect to the $x$-axis of the source frame. Due to precession, spins
measurements vary between the two reference points but can be related by a spin
evolution~\cite{Varma:2019csw, Varma:2021csh}.

For both reference points in Fig.~\ref{fig:corner_GW200129}, there is a clear
preference for large orbital-plane spins for the heavier BH (large $\chi_1$ and
$\cos \theta_1 \sim 0$). Even though the spin of the lighter BH is not well
measured, this is sufficient for precession. We stress that while precessing
binaries tend to have larger kicks~\cite{Campanelli:2007cga, Gonzalez:2007hi,
Lousto:2011kp}, precession does not necessarily imply a large kick, and it is
important to directly compute the kick velocity as we do in the next section.
In particular, the kick can vary from zero to $\sim 5000$ km/s just by changing
the azimuthal spin angles, even for systems with large orbital-plane
spins~\cite{Campanelli:2007cga, Gonzalez:2007hi, Lousto:2011kp}.

Next, the azimuthal angles (especially $\phi_1$) in
Fig.~\ref{fig:corner_GW200129} are much better constrained at $\trefmHundredM$,
while the other parameters do not change significantly.~\footnote{The
    posteriors for $\chi_{1}$, $\chi_{2}$, and $q$ are expected to be
consistent between the two reference points (modulo parameter estimation
uncertainty) as these parameters are independent of the reference point.} This
feature is key: even though the azimuthal angles are poorly constrained in the
inspiral, they are well constrained at $\trefmHundredM$~\cite{Varma:2021csh}.
As the kick depends sensitively on the azimuthal angles near the
merger~\cite{Campanelli:2007cga}, successfully measuring these angles at
$\trefmHundredM$ is critical for constraining the kick.

Spins measured at $\frefTwentyHz$ can also be evolved consistently to
$\trefmHundredM$ using \NRSur dynamics~\cite{Varma:2019csw}. In fact, this
procedure is internally applied by \NRSurRemnant if the spins are specified at
$\frefTwentyHz$~\cite{Varma:2019csw, Varma:2018aht}. Therefore, by
construction, the kick posterior for individual GW events is independent of the
reference point at which the spins are initially measured (modulo \NRSur spin
evolution errors, which are small compared to the model
errors~\cite{Varma:2019csw, Blackman:2017pcm}). Therefore, for the purpose of
this paper, the main benefit of the spin measurements at $\trefmHundredM$ is to
illustrate why a successful kick constraint is possible in the first place. The
supplement of Ref.~\cite{Varma:2021xbh} discusses other benefits, in
particular, for constraining the ensemble population of spins and kicks. In the
rest of the paper, we will use the spin measurements at $\trefmHundredM$.

As noted in Refs.~\cite{LIGOScientific:2021djp, Hannam:2021pit}, the inference
of precession in GW200129 depends on the waveform model used. In particular,
while the phenomenological model \Phenom~\cite{Pratten:2020ceb} recovers
precession, the effective-one-body model \EOB~\cite{Ossokine:2020kjp} does not.
Among these models, only \NRSur is informed by precessing NR simulations and is
more accurate by about an order of magnitude~\cite{Varma:2019csw}. By contrast,
\EOB and \Phenom approximate precession effects by ``twisting'' the frame of an
equivalent aligned-spin binary~\cite{Ossokine:2020kjp, Pratten:2020ceb}.
Furthermore, Ref.~\cite{Varma:2021csh} found that \NRSur is necessary to
accurately measure the spin vectors $\bchi_{1,2}$, in particular, the spin
directions within the orbital plane~\cite{Varma:2021csh}, which have a strong
influence on the kick~\cite{Campanelli:2007cga}. Similarly, given the spin
measurements, \NRSurRemnant is necessary to accurately predict the kick
velocity~\cite{Varma:2020nbm}. For these reasons, we treat \NRSur and
\NRSurRemnant as the preferred models for analyzing GW200129.

\prlsec{GW200129 kick velocity.}
Figure~\ref{fig:kick} shows our constraints on the kick magnitude $v_f$ of
GW200129, obtained by evaluating \NRSurRemnant on the \NRSur $\bLambda$
posteriors at $\trefmHundredM$. In the left panel, we show the posterior and
prior distributions for $v_f$, along with fiducial escape velocities for
globular clusters~\cite{Antonini:2016gqe}, nuclear star
clusters~\cite{Antonini:2016gqe}, giant elliptical
galaxies~\cite{Merritt:2004xa} and Milky Way-like
galaxies~\cite{Monari:2018esc} for comparison. Unlike the events considered in
Ref.~\cite{Varma:2020nbm}, the $v_f$ posterior is clearly distinguishable from
the prior, and there is substantial information gain about the kick.\footnote{
The Kullback--Leibler (KL) divergence~\cite{Kullback_Leibler_divergence} from
the prior to the posterior in Fig.~\ref{fig:kick} is \KLDiv bits. By contrast,
the largest KL divergence for the events considered in
Ref.~\cite{Varma:2020nbm} was 0.22 bits.
}

The kick magnitude is constrained to \vRange (median and 90\% symmetric
credible interval), or \vLower (lower 10th percentile), making GW200129 the
first GW event identified as having a large kick velocity.  We note, however,
that such large kick velocities are not surprising given previous constraints
on the ensemble properties of merging binary BHs~\cite{Varma:2021xbh,
Doctor:2021qfn}. For example, Fig.~3 of Ref.~\cite{Varma:2021xbh}, which shows
estimates of the ensemble kick distribution, includes nonneglibile support up
to $v_f \sim 1500$ km/s.

The large kick of GW200129 raises the question of whether the remnant BH is
ejected from its host environment. This has implications for the formation of
heavy BHs through second-generation mergers in dense
environments~\cite{Gerosa:2021mno}. This formation channel is one possible way
to explain observations of BHs with masses $\gtrsim 65
M_{\odot}$~\cite{Abbott:2020tfl, Abbott:2020niy, LIGOScientific:2021djp}, which
fall within the mass gap expected due to the (pulsational) pair-instability
supernova processes~\cite{Woosley:2016hmi, Marchant:2018kun}.  To address this,
we compute the retention probability for the remnant BH of GW200129 in globular
clusters and nuclear star clusters, both of which host dense stellar
environments where merger remnants can potentially interact with other BHs and
form binaries.

The right panel of Fig.~\ref{fig:kick} shows the cumulative distribution
functions (CDFs) for the $v_f$ posterior and prior. As the posterior CDF($v_f$)
denotes the probability that the kick magnitude of GW200129 is below $v_f$, we
take it to be the probability that the remnant BH is retained by a host
environment with an escape velocity of $v_f$. The vertical dotted lines
indicate the maximum escape velocity $\vesc$ for various host environments;
CDF($\vesc$) sets the upper limit on the retention probability for that host.
In particular, assuming $\vesc\!=\!100$ km/s ($\vesc\!=\!600$) for globular
(nuclear star) clusters there is a less than $0.48\%$ ($7.7\%$) probability
that the remnant BH of GW200129 is retained by those hosts. This is consistent
with Refs.~\cite{Varma:2021xbh, Doctor:2021qfn}, where globular clusters were
already identified as an unlikely site for second generation mergers, even for
more moderate kicks.

\begin{figure}[thb]
\includegraphics[width=0.45\textwidth]{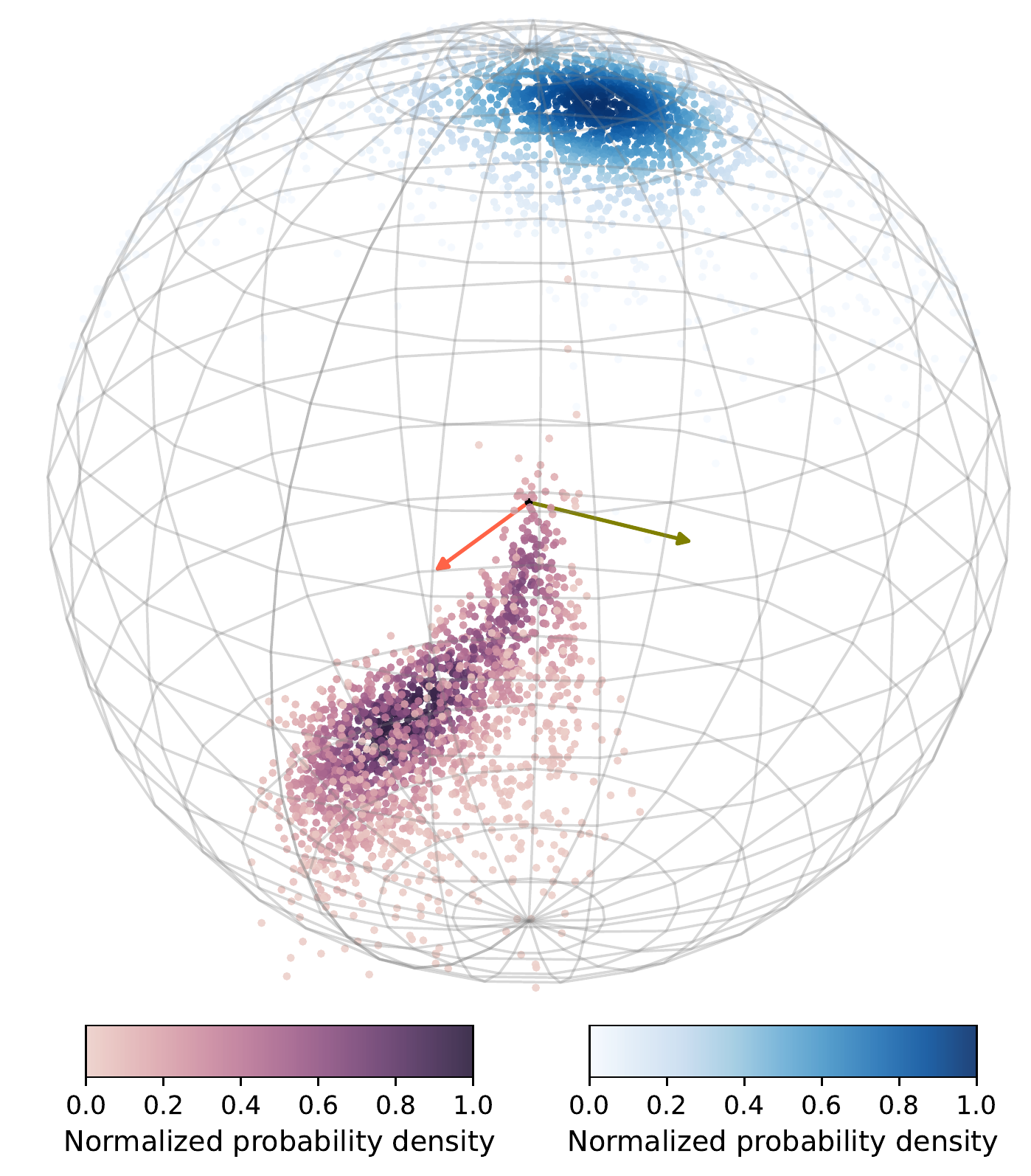}
\caption{
Posterior samples for the full kick vector $\bv_f$ in the source frame at
$\trefmHundredM$. Each purple marker indicates a kick posterior sample; an
arrow drawn from the origin to the marker would show the kick vector $\bv_f$.
The outer radius of the sphere corresponds to $v_f=2500$ km/s. The $x$-axis
(orange) and $y$-axis (green) are shown as arrows near the origin; the $x-y$
plane is orthogonal to the orbital angular momentum direction. The blue markers
on the sphere show posterior samples for the line-of-sight direction to the
observer. For both distributions, the spread represents the measurement
uncertainty, and the color reflects posterior probability density (normalized
so that the peak density is 1). A rotating perspective of this plot can be seen
at
\href{https://vijayvarma392.github.io/GW200129/\#kick}
{vijayvarma392.github.io/GW200129/\#kick}.
}
\label{fig:kick_direction}
\end{figure}

\prlsec{Remnant mass and Doppler shifts.}
Our method provides predictions for both the magnitude and direction of the
kick~\cite{Varma:2020nbm}. If the kick vector $\bv_f$ has a significant
component along (or opposite) the line-of-sight, the observed GW signal can be
influenced by the kick.  At leading order, the kick's effect can be described
as a Doppler shift of the GW frequency~\cite{Gerosa:2016vip}. However, as GR
lacks any intrinsic length scales, a uniform increase in signal frequency is
completely degenerate with a decrease in total mass $M$, and vice versa. Thus,
if not explicitly accounted for, a frequency shift due to a kick can bias mass
measurements. In particular, because the kick is mostly imparted near the
merger~\cite{Gonzalez:2006md}, the Doppler shift only affects the merger and
ringdown part of the signal. This can lead to biases in the measurement of the
remnant mass $m_f$~\cite{Varma:2020nbm, Ma:2021znq}, and potentially impact
tests of GR using the ringdown signal~\cite{LIGOScientific:2021sio}. However,
this effect is expected to be small for current detectors~\cite{Varma:2020nbm,
Gerosa:2016vip}.

\begin{figure}[thb]
\includegraphics[width=0.45\textwidth]{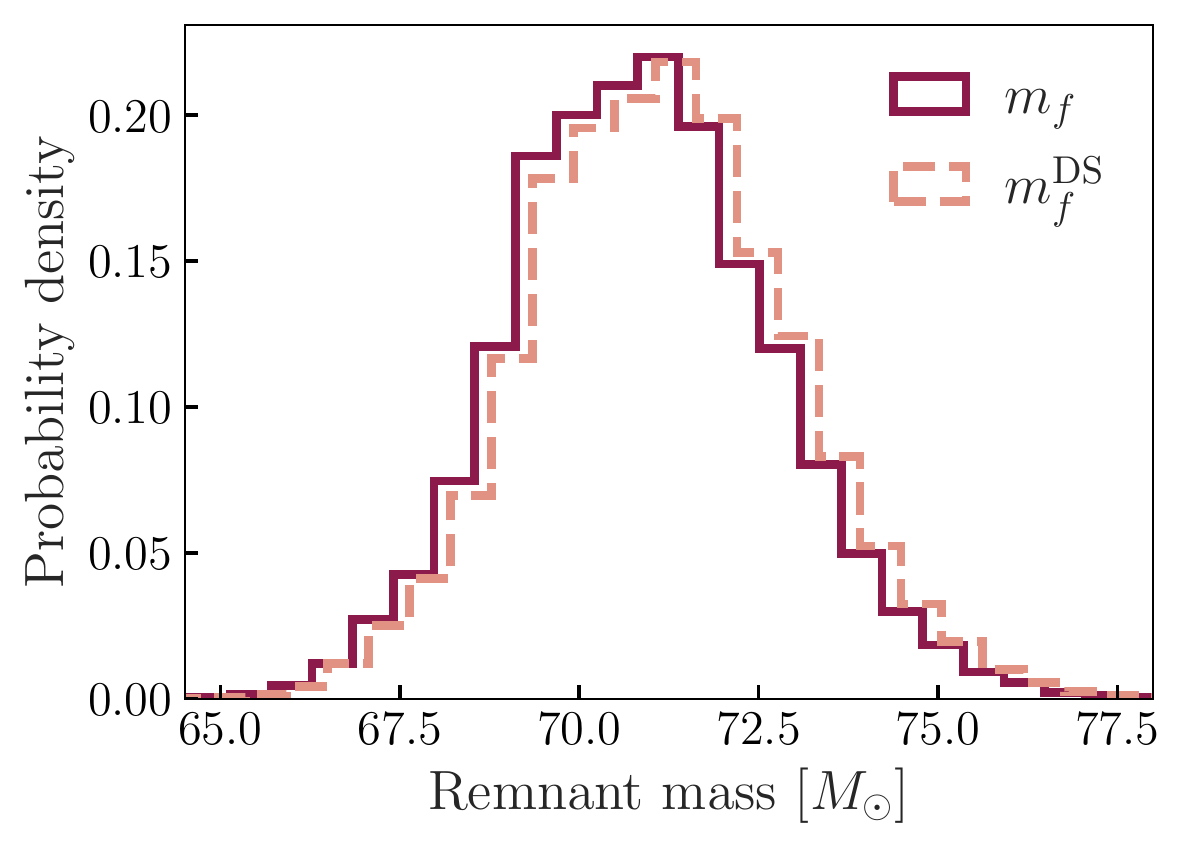}
\caption{
The remnant mass and the Doppler shifted remnant mass for GW200129, as inferred
in the detector frame. There is an overall redshift, as the kick direction in
Fig.~\ref{fig:kick_direction} is pointed (roughly) away from the observer.
However, as these distributions are very close, we do not expect ringdown tests
of GR to be impacted by the kick for this event.
}
\label{fig:mf_doppler}
\end{figure}

In the following, we verify that the Doppler effect on the remnant mass
measurement of GW200129 is indeed small. At leading order, the Doppler-shifted
remnant mass is given by~\cite{Gerosa:2016vip}:
\begin{equation}
m^{\text{DS}}_f = m_f\,(1 + \bv_f \cdot \hat{\bm{n}}/c),
\label{eq:doppler_mass}
\end{equation}
where $c$ is the speed of light and $\hat{\bm{n}}$ is the unit vector pointing
along the line-of-sight from the observer to the source. The line-of-sight
direction is obtained from our inference setup, parameterized by ($\iota$,
$\phi$). $\iota$ is the inclination angle between the orbital angular momentum
and the line-of-sight to the observer, and $\phi$ is the azimuthal angle to the
observer in the orbital plane, both defined in the source frame at
$\trefmHundredM$.

Figure~\ref{fig:kick_direction} shows the posterior distributions for the full
kick vector $\bv_f$ (also defined in the source frame at $\trefmHundredM$) and
the line-of-sight direction. We find that the kick and the line-of-sight are
not very well (anti-) aligned; therefore, we do not expect significant Doppler
shifts for this signal. Finally, Fig.~\ref{fig:mf_doppler} shows the posterior
distributions for $m_f$ (obtained from \NRSurRemnant) and $m^{\text{DS}}_f$
(computed using Eq.~(\ref{eq:doppler_mass})) for GW200129. As expected, the
difference between these distributions is very small compared to the
measurement uncertainty, meaning that tests of GR should not be impacted by
the Doppler effect for this event.  As detector sensitivity improves, this may
not be the case, however, and it may be necessary to explicitly account for
this effect~\cite{Varma:2020nbm}.

\prlsec{Conclusions.}
We use NR surrogate models for the gravitational waveform and the remnant BH
properties to infer the kick velocity for the binary BH merger GW200129. The
kick magnitude is constrained to \vRange or \vLower, at 90\%
credibility. Given the kick velocity, we estimate that there is at most a
$0.48\%$ ($7.7\%$) probability that the remnant BH of GW200129 would be
retained by globular (nuclear star) clusters.  Finally, we show that the
Doppler effect on the remnant mass is small compared to current measurement
uncertainty; therefore ringdown tests of GR are not expected to be
significantly impacted by the kick for this event.

Observational evidence for kicks has far reaching implications for BH
astrophysics. GW200129 is the first GW event identified as having a large kick
velocity. Large kicks like this have been previously predicted based on the
ensemble kick distribution of merging binary BHs~\cite{Varma:2021xbh,
Doctor:2021qfn}, and we can expect to see more such events as detector
sensitivity improves. In particular, such observations can help resolve the
mystery of the heavy BHs seen by LIGO-Virgo~\cite{Abbott:2020tfl,
Abbott:2020niy, LIGOScientific:2021djp}, by constraining the rate of
second-generation mergers.

\vspace{0.5cm}
\prlsec{Acknowledgments.}
We thank Arif Shaikh for comments on the manuscript.
V.V acknowledges funding from the European Union’s Horizon 2020 research and
innovation program under the Marie Skłodowska-Curie grant agreement No.~896869.
S.B., C.-J.H., and S.V. acknowledge support of the National Science Foundation
(NSF) and the LIGO Laboratory. S.B. is also supported by the NSF Graduate
Research Fellowship under Grant No.~DGE-1122374. S.V. is also supported by NSF
Grant No.~PHY-2045740.
T.I. is supported by the Heising-Simons Foundation, the Simons Foundation, and
NSF Grants Nos.~PHY-1748958, PHY-1806665 and DMS-1912716. F.S. and S.E.F. are
supported by NSF Grants Nos.~PHY-2110496 and PHY-1806665.
Computations were performed on the Wheeler cluster at Caltech, which is
supported by the Sherman Fairchild Foundation and by Caltech; and the High
Performance Cluster at Caltech.
This material is based upon work supported by NSF's LIGO Laboratory which is a
major facility fully funded by the NSF. LIGO was constructed by the California
Institute of Technology and Massachusetts Institute of Technology with funding
from the NSF and operates under cooperative agreement PHY-0757058.
This research made use of data, software and/or web tools obtained from the
Gravitational Wave Open Science Center~\cite{GW_open_science_center}, a service
of the LIGO Laboratory, the LIGO Scientific Collaboration and the Virgo
Collaboration.

\bibliography{References}

\begin{thebibliography}{62}%
\makeatletter
\providecommand \@ifxundefined [1]{%
 \@ifx{#1\undefined}
}%
\providecommand \@ifnum [1]{%
 \ifnum #1\expandafter \@firstoftwo
 \else \expandafter \@secondoftwo
 \fi
}%
\providecommand \@ifx [1]{%
 \ifx #1\expandafter \@firstoftwo
 \else \expandafter \@secondoftwo
 \fi
}%
\providecommand \natexlab [1]{#1}%
\providecommand \enquote  [1]{``#1''}%
\providecommand \bibnamefont  [1]{#1}%
\providecommand \bibfnamefont [1]{#1}%
\providecommand \citenamefont [1]{#1}%
\providecommand \href@noop [0]{\@secondoftwo}%
\providecommand \href [0]{\begingroup \@sanitize@url \@href}%
\providecommand \@href[1]{\@@startlink{#1}\@@href}%
\providecommand \@@href[1]{\endgroup#1\@@endlink}%
\providecommand \@sanitize@url [0]{\catcode `\\12\catcode `\$12\catcode
  `\&12\catcode `\#12\catcode `\^12\catcode `\_12\catcode `\%12\relax}%
\providecommand \@@startlink[1]{}%
\providecommand \@@endlink[0]{}%
\providecommand \url  [0]{\begingroup\@sanitize@url \@url }%
\providecommand \@url [1]{\endgroup\@href {#1}{\urlprefix }}%
\providecommand \urlprefix  [0]{URL }%
\providecommand \Eprint [0]{\href }%
\providecommand \doibase [0]{http://dx.doi.org/}%
\providecommand \selectlanguage [0]{\@gobble}%
\providecommand \bibinfo  [0]{\@secondoftwo}%
\providecommand \bibfield  [0]{\@secondoftwo}%
\providecommand \translation [1]{[#1]}%
\providecommand \BibitemOpen [0]{}%
\providecommand \bibitemStop [0]{}%
\providecommand \bibitemNoStop [0]{.\EOS\space}%
\providecommand \EOS [0]{\spacefactor3000\relax}%
\providecommand \BibitemShut  [1]{\csname bibitem#1\endcsname}%
\let\auto@bib@innerbib\@empty
\bibitem [{\citenamefont {Fitchett}(1983)}]{Fitchett:1983MNRAS}%
  \BibitemOpen
  \bibfield  {author} {\bibinfo {author} {\bibfnamefont {M.~J.}\ \bibnamefont
  {Fitchett}},\ }\bibfield  {title} {\enquote {\bibinfo {title} {{The influence
  of gravitational wave momentum losses on the centre of mass motion of a
  Newtonian binary system}},}\ }\href {\doibase 10.1093/mnras/203.4.1049}
  {\bibfield  {journal} {\bibinfo  {journal} {Monthly Notices of the Royal
  Astronomical Society}\ }\textbf {\bibinfo {volume} {203}},\ \bibinfo {pages}
  {1049--1062} (\bibinfo {year} {1983})},\ \Eprint
  {http://arxiv.org/abs/http://oup.prod.sis.lan/mnras/article-pdf/203/4/1049/18223796/mnras203-1049.pdf}
  {http://oup.prod.sis.lan/mnras/article-pdf/203/4/1049/18223796/mnras203-1049.pdf}
  \BibitemShut {NoStop}%
\bibitem [{\citenamefont {Gonzalez}\ \emph
  {et~al.}(2007{\natexlab{a}})\citenamefont {Gonzalez}, \citenamefont
  {Sperhake}, \citenamefont {Bruegmann}, \citenamefont {Hannam},\ and\
  \citenamefont {Husa}}]{Gonzalez:2006md}%
  \BibitemOpen
  \bibfield  {author} {\bibinfo {author} {\bibfnamefont {Jose~A.}\ \bibnamefont
  {Gonzalez}}, \bibinfo {author} {\bibfnamefont {Ulrich}\ \bibnamefont
  {Sperhake}}, \bibinfo {author} {\bibfnamefont {Bernd}\ \bibnamefont
  {Bruegmann}}, \bibinfo {author} {\bibfnamefont {Mark}\ \bibnamefont
  {Hannam}}, \ and\ \bibinfo {author} {\bibfnamefont {Sascha}\ \bibnamefont
  {Husa}},\ }\bibfield  {title} {\enquote {\bibinfo {title} {{Total recoil: The
  Maximum kick from nonspinning black-hole binary inspiral}},}\ }\href
  {\doibase 10.1103/PhysRevLett.98.091101} {\bibfield  {journal} {\bibinfo
  {journal} {Phys. Rev. Lett.}\ }\textbf {\bibinfo {volume} {98}},\ \bibinfo
  {pages} {091101} (\bibinfo {year} {2007}{\natexlab{a}})},\ \Eprint
  {http://arxiv.org/abs/gr-qc/0610154} {arXiv:gr-qc/0610154 [gr-qc]}
  \BibitemShut {NoStop}%
\bibitem [{\citenamefont {Apostolatos}\ \emph {et~al.}(1994)\citenamefont
  {Apostolatos}, \citenamefont {Cutler}, \citenamefont {Sussman},\ and\
  \citenamefont {Thorne}}]{Apostolatos:1994pre}%
  \BibitemOpen
  \bibfield  {author} {\bibinfo {author} {\bibfnamefont {Theocharis~A.}\
  \bibnamefont {Apostolatos}}, \bibinfo {author} {\bibfnamefont {Curt}\
  \bibnamefont {Cutler}}, \bibinfo {author} {\bibfnamefont {Gerald~J.}\
  \bibnamefont {Sussman}}, \ and\ \bibinfo {author} {\bibfnamefont {Kip~S.}\
  \bibnamefont {Thorne}},\ }\bibfield  {title} {\enquote {\bibinfo {title}
  {Spin-induced orbital precession and its modulation of the gravitational
  waveforms from merging binaries},}\ }\href {\doibase
  10.1103/PhysRevD.49.6274} {\bibfield  {journal} {\bibinfo  {journal} {Phys.
  Rev. D}\ }\textbf {\bibinfo {volume} {49}},\ \bibinfo {pages} {6274--6297}
  (\bibinfo {year} {1994})}\BibitemShut {NoStop}%
\bibitem [{\citenamefont {Campanelli}\ \emph {et~al.}(2007)\citenamefont
  {Campanelli}, \citenamefont {Lousto}, \citenamefont {Zlochower},\ and\
  \citenamefont {Merritt}}]{Campanelli:2007cga}%
  \BibitemOpen
  \bibfield  {author} {\bibinfo {author} {\bibfnamefont {Manuela}\ \bibnamefont
  {Campanelli}}, \bibinfo {author} {\bibfnamefont {Carlos~O.}\ \bibnamefont
  {Lousto}}, \bibinfo {author} {\bibfnamefont {Yosef}\ \bibnamefont
  {Zlochower}}, \ and\ \bibinfo {author} {\bibfnamefont {David}\ \bibnamefont
  {Merritt}},\ }\bibfield  {title} {\enquote {\bibinfo {title} {{Maximum
  gravitational recoil}},}\ }\href {\doibase 10.1103/PhysRevLett.98.231102}
  {\bibfield  {journal} {\bibinfo  {journal} {Phys. Rev. Lett.}\ }\textbf
  {\bibinfo {volume} {98}},\ \bibinfo {pages} {231102} (\bibinfo {year}
  {2007})},\ \Eprint {http://arxiv.org/abs/gr-qc/0702133} {arXiv:gr-qc/0702133
  [GR-QC]} \BibitemShut {NoStop}%
\bibitem [{\citenamefont {Gonzalez}\ \emph
  {et~al.}(2007{\natexlab{b}})\citenamefont {Gonzalez}, \citenamefont {Hannam},
  \citenamefont {Sperhake}, \citenamefont {Bruegmann},\ and\ \citenamefont
  {Husa}}]{Gonzalez:2007hi}%
  \BibitemOpen
  \bibfield  {author} {\bibinfo {author} {\bibfnamefont {J.~A.}\ \bibnamefont
  {Gonzalez}}, \bibinfo {author} {\bibfnamefont {M.~D.}\ \bibnamefont
  {Hannam}}, \bibinfo {author} {\bibfnamefont {U.}~\bibnamefont {Sperhake}},
  \bibinfo {author} {\bibfnamefont {Bernd}\ \bibnamefont {Bruegmann}}, \ and\
  \bibinfo {author} {\bibfnamefont {S.}~\bibnamefont {Husa}},\ }\bibfield
  {title} {\enquote {\bibinfo {title} {{Supermassive recoil velocities for
  binary black-hole mergers with antialigned spins}},}\ }\href {\doibase
  10.1103/PhysRevLett.98.231101} {\bibfield  {journal} {\bibinfo  {journal}
  {Phys. Rev. Lett.}\ }\textbf {\bibinfo {volume} {98}},\ \bibinfo {pages}
  {231101} (\bibinfo {year} {2007}{\natexlab{b}})},\ \Eprint
  {http://arxiv.org/abs/gr-qc/0702052} {arXiv:gr-qc/0702052 [GR-QC]}
  \BibitemShut {NoStop}%
\bibitem [{\citenamefont {Lousto}\ and\ \citenamefont
  {Zlochower}(2011)}]{Lousto:2011kp}%
  \BibitemOpen
  \bibfield  {author} {\bibinfo {author} {\bibfnamefont {Carlos~O.}\
  \bibnamefont {Lousto}}\ and\ \bibinfo {author} {\bibfnamefont {Yosef}\
  \bibnamefont {Zlochower}},\ }\bibfield  {title} {\enquote {\bibinfo {title}
  {{Hangup Kicks: Still Larger Recoils by Partial Spin/Orbit Alignment of
  Black-Hole Binaries}},}\ }\href {\doibase 10.1103/PhysRevLett.107.231102}
  {\bibfield  {journal} {\bibinfo  {journal} {Phys. Rev. Lett.}\ }\textbf
  {\bibinfo {volume} {107}},\ \bibinfo {pages} {231102} (\bibinfo {year}
  {2011})},\ \Eprint {http://arxiv.org/abs/1108.2009} {arXiv:1108.2009 [gr-qc]}
  \BibitemShut {NoStop}%
\bibitem [{\citenamefont {Merritt}\ \emph {et~al.}(2004)\citenamefont
  {Merritt}, \citenamefont {Milosavljevic}, \citenamefont {Favata},
  \citenamefont {Hughes},\ and\ \citenamefont {Holz}}]{Merritt:2004xa}%
  \BibitemOpen
  \bibfield  {author} {\bibinfo {author} {\bibfnamefont {David}\ \bibnamefont
  {Merritt}}, \bibinfo {author} {\bibfnamefont {Milos}\ \bibnamefont
  {Milosavljevic}}, \bibinfo {author} {\bibfnamefont {Marc}\ \bibnamefont
  {Favata}}, \bibinfo {author} {\bibfnamefont {Scott~A.}\ \bibnamefont
  {Hughes}}, \ and\ \bibinfo {author} {\bibfnamefont {Daniel~E.}\ \bibnamefont
  {Holz}},\ }\bibfield  {title} {\enquote {\bibinfo {title} {{Consequences of
  gravitational radiation recoil}},}\ }\href {\doibase 10.1086/421551}
  {\bibfield  {journal} {\bibinfo  {journal} {Astrophys. J.}\ }\textbf
  {\bibinfo {volume} {607}},\ \bibinfo {pages} {L9--L12} (\bibinfo {year}
  {2004})},\ \Eprint {http://arxiv.org/abs/astro-ph/0402057}
  {arXiv:astro-ph/0402057 [astro-ph]} \BibitemShut {NoStop}%
\bibitem [{\citenamefont {Komossa}\ and\ \citenamefont
  {Merritt}(2008)}]{Komossa:2008as}%
  \BibitemOpen
  \bibfield  {author} {\bibinfo {author} {\bibfnamefont {S.}~\bibnamefont
  {Komossa}}\ and\ \bibinfo {author} {\bibfnamefont {David}\ \bibnamefont
  {Merritt}},\ }\bibfield  {title} {\enquote {\bibinfo {title} {{Gravitational
  Wave Recoil Oscillations of Black Holes: Implications for Unified Models of
  Active Galactic Nuclei}},}\ }\href {\doibase 10.1086/595883} {\bibfield
  {journal} {\bibinfo  {journal} {Astrophys. J.}\ }\textbf {\bibinfo {volume}
  {689}},\ \bibinfo {pages} {L89} (\bibinfo {year} {2008})},\ \Eprint
  {http://arxiv.org/abs/0811.1037} {arXiv:0811.1037 [astro-ph]} \BibitemShut
  {NoStop}%
\bibitem [{\citenamefont {{Volonteri}}\ \emph {et~al.}(2010)\citenamefont
  {{Volonteri}}, \citenamefont {{G{\"u}ltekin}},\ and\ \citenamefont
  {{Dotti}}}]{Volonteri:2010mbh}%
  \BibitemOpen
  \bibfield  {author} {\bibinfo {author} {\bibfnamefont {Marta}\ \bibnamefont
  {{Volonteri}}}, \bibinfo {author} {\bibfnamefont {Kayhan}\ \bibnamefont
  {{G{\"u}ltekin}}}, \ and\ \bibinfo {author} {\bibfnamefont {Massimo}\
  \bibnamefont {{Dotti}}},\ }\bibfield  {title} {\enquote {\bibinfo {title}
  {{Gravitational recoil: effects on massive black hole occupation fraction
  over cosmic time}},}\ }\href {\doibase 10.1111/j.1365-2966.2010.16431.x}
  {\bibfield  {journal} {\bibinfo  {journal} {Monthly Notices of the Royal
  Astronomical Society}\ }\textbf {\bibinfo {volume} {404}},\ \bibinfo {pages}
  {2143--2150} (\bibinfo {year} {2010})},\ \Eprint
  {http://arxiv.org/abs/1001.1743} {arXiv:1001.1743 [astro-ph.CO]} \BibitemShut
  {NoStop}%
\bibitem [{\citenamefont {Sesana}(2007)}]{sesana:2007zk}%
  \BibitemOpen
  \bibfield  {author} {\bibinfo {author} {\bibfnamefont {A.}~\bibnamefont
  {Sesana}},\ }\bibfield  {title} {\enquote {\bibinfo {title} {{Extreme
  recoils: impact on the detection of gravitational waves from massive black
  hole binaries}},}\ }\href {\doibase 10.1111/j.1745-3933.2007.00375.x}
  {\bibfield  {journal} {\bibinfo  {journal} {Mon. Not. Roy. Astron. Soc.}\
  }\textbf {\bibinfo {volume} {382}},\ \bibinfo {pages} {6} (\bibinfo {year}
  {2007})},\ \Eprint {http://arxiv.org/abs/0707.4677} {arXiv:0707.4677
  [astro-ph]} \BibitemShut {NoStop}%
\bibitem [{\citenamefont {Amaro-Seoane}\ \emph {et~al.}(2017)\citenamefont
  {Amaro-Seoane} \emph {et~al.}}]{AmaroSeoane:2017las}%
  \BibitemOpen
  \bibfield  {author} {\bibinfo {author} {\bibfnamefont {Pau}\ \bibnamefont
  {Amaro-Seoane}} \emph {et~al.},\ }\bibfield  {title} {\enquote {\bibinfo
  {title} {{Laser Interferometer Space Antenna}},}\ }\href@noop {} {\bibfield
  {journal} {\bibinfo  {journal} {arXiv e-prints}\ ,\ \bibinfo {eid}
  {arXiv:1702.00786}} (\bibinfo {year} {2017})},\ \Eprint
  {http://arxiv.org/abs/1702.00786} {arXiv:1702.00786 [astro-ph.IM]}
  \BibitemShut {NoStop}%
\bibitem [{\citenamefont {Aasi}\ \emph {et~al.}(2015)\citenamefont {Aasi} \emph
  {et~al.}}]{TheLIGOScientific:2014jea}%
  \BibitemOpen
  \bibfield  {author} {\bibinfo {author} {\bibfnamefont {J.}~\bibnamefont
  {Aasi}} \emph {et~al.} (\bibinfo {collaboration} {LIGO Scientific}),\
  }\bibfield  {title} {\enquote {\bibinfo {title} {{Advanced LIGO}},}\ }\href
  {\doibase 10.1088/0264-9381/32/7/074001} {\bibfield  {journal} {\bibinfo
  {journal} {Class. Quant. Grav.}\ }\textbf {\bibinfo {volume} {32}},\ \bibinfo
  {pages} {074001} (\bibinfo {year} {2015})},\ \Eprint
  {http://arxiv.org/abs/1411.4547} {arXiv:1411.4547 [gr-qc]} \BibitemShut
  {NoStop}%
\bibitem [{\citenamefont {Acernese}\ \emph {et~al.}(2015)\citenamefont
  {Acernese} \emph {et~al.}}]{TheVirgo:2014hva}%
  \BibitemOpen
  \bibfield  {author} {\bibinfo {author} {\bibfnamefont {F.}~\bibnamefont
  {Acernese}} \emph {et~al.} (\bibinfo {collaboration} {Virgo}),\ }\bibfield
  {title} {\enquote {\bibinfo {title} {{Advanced Virgo: a second-generation
  interferometric gravitational wave detector}},}\ }\href {\doibase
  10.1088/0264-9381/32/2/024001} {\bibfield  {journal} {\bibinfo  {journal}
  {Class. Quant. Grav.}\ }\textbf {\bibinfo {volume} {32}},\ \bibinfo {pages}
  {024001} (\bibinfo {year} {2015})},\ \Eprint {http://arxiv.org/abs/1408.3978}
  {arXiv:1408.3978 [gr-qc]} \BibitemShut {NoStop}%
\bibitem [{\citenamefont {Woosley}(2017)}]{Woosley:2016hmi}%
  \BibitemOpen
  \bibfield  {author} {\bibinfo {author} {\bibfnamefont {S.~E.}\ \bibnamefont
  {Woosley}},\ }\bibfield  {title} {\enquote {\bibinfo {title} {{Pulsational
  Pair-Instability Supernovae}},}\ }\href {\doibase
  10.3847/1538-4357/836/2/244} {\bibfield  {journal} {\bibinfo  {journal}
  {Astrophys. J.}\ }\textbf {\bibinfo {volume} {836}},\ \bibinfo {pages} {244}
  (\bibinfo {year} {2017})},\ \Eprint {http://arxiv.org/abs/1608.08939}
  {arXiv:1608.08939 [astro-ph.HE]} \BibitemShut {NoStop}%
\bibitem [{\citenamefont {Marchant}\ \emph {et~al.}(2018)\citenamefont
  {Marchant}, \citenamefont {Renzo}, \citenamefont {Farmer}, \citenamefont
  {Pappas}, \citenamefont {Taam}, \citenamefont {de~Mink},\ and\ \citenamefont
  {Kalogera}}]{Marchant:2018kun}%
  \BibitemOpen
  \bibfield  {author} {\bibinfo {author} {\bibfnamefont {Pablo}\ \bibnamefont
  {Marchant}}, \bibinfo {author} {\bibfnamefont {Mathieu}\ \bibnamefont
  {Renzo}}, \bibinfo {author} {\bibfnamefont {Robert}\ \bibnamefont {Farmer}},
  \bibinfo {author} {\bibfnamefont {Kaliroe M.~W.}\ \bibnamefont {Pappas}},
  \bibinfo {author} {\bibfnamefont {Ronald~E.}\ \bibnamefont {Taam}}, \bibinfo
  {author} {\bibfnamefont {Selma}\ \bibnamefont {de~Mink}}, \ and\ \bibinfo
  {author} {\bibfnamefont {Vassiliki}\ \bibnamefont {Kalogera}},\ }\bibfield
  {title} {\enquote {\bibinfo {title} {{Pulsational pair-instability supernovae
  in very close binaries}},}\ }\href {\doibase 10.3847/1538-4357/ab3426} {\
  (\bibinfo {year} {2018}),\ 10.3847/1538-4357/ab3426},\ \Eprint
  {http://arxiv.org/abs/1810.13412} {arXiv:1810.13412 [astro-ph.HE]}
  \BibitemShut {NoStop}%
\bibitem [{\citenamefont {Abbott}\ \emph
  {et~al.}(2020{\natexlab{a}})\citenamefont {Abbott} \emph
  {et~al.}}]{Abbott:2020tfl}%
  \BibitemOpen
  \bibfield  {author} {\bibinfo {author} {\bibfnamefont {R.}~\bibnamefont
  {Abbott}} \emph {et~al.} (\bibinfo {collaboration} {LIGO Scientific,
  Virgo}),\ }\bibfield  {title} {\enquote {\bibinfo {title} {{GW190521: A
  Binary Black Hole Merger with a Total Mass of $150 ~ M_{\odot}$}},}\ }\href
  {\doibase 10.1103/PhysRevLett.125.101102} {\bibfield  {journal} {\bibinfo
  {journal} {Phys. Rev. Lett.}\ }\textbf {\bibinfo {volume} {125}},\ \bibinfo
  {pages} {101102} (\bibinfo {year} {2020}{\natexlab{a}})},\ \Eprint
  {http://arxiv.org/abs/2009.01075} {arXiv:2009.01075 [gr-qc]} \BibitemShut
  {NoStop}%
\bibitem [{\citenamefont {Abbott}\ \emph
  {et~al.}(2021{\natexlab{a}})\citenamefont {Abbott} \emph
  {et~al.}}]{Abbott:2020niy}%
  \BibitemOpen
  \bibfield  {author} {\bibinfo {author} {\bibfnamefont {R.}~\bibnamefont
  {Abbott}} \emph {et~al.} (\bibinfo {collaboration} {LIGO Scientific,
  Virgo}),\ }\bibfield  {title} {\enquote {\bibinfo {title} {{GWTC-2: Compact
  Binary Coalescences Observed by LIGO and Virgo During the First Half of the
  Third Observing Run}},}\ }\href {\doibase 10.1103/PhysRevX.11.021053}
  {\bibfield  {journal} {\bibinfo  {journal} {Phys. Rev. X}\ }\textbf {\bibinfo
  {volume} {11}},\ \bibinfo {pages} {021053} (\bibinfo {year}
  {2021}{\natexlab{a}})},\ \Eprint {http://arxiv.org/abs/2010.14527}
  {arXiv:2010.14527 [gr-qc]} \BibitemShut {NoStop}%
\bibitem [{\citenamefont {Abbott}\ \emph
  {et~al.}(2021{\natexlab{b}})\citenamefont {Abbott} \emph
  {et~al.}}]{LIGOScientific:2021djp}%
  \BibitemOpen
  \bibfield  {author} {\bibinfo {author} {\bibfnamefont {R.}~\bibnamefont
  {Abbott}} \emph {et~al.} (\bibinfo {collaboration} {LIGO Scientific, VIRGO,
  KAGRA}),\ }\bibfield  {title} {\enquote {\bibinfo {title} {{GWTC-3: Compact
  Binary Coalescences Observed by LIGO and Virgo During the Second Part of the
  Third Observing Run}},}\ }\href@noop {} {\  (\bibinfo {year}
  {2021}{\natexlab{b}})},\ \Eprint {http://arxiv.org/abs/2111.03606}
  {arXiv:2111.03606 [gr-qc]} \BibitemShut {NoStop}%
\bibitem [{\citenamefont {Gerosa}\ and\ \citenamefont
  {Fishbach}(2021)}]{Gerosa:2021mno}%
  \BibitemOpen
  \bibfield  {author} {\bibinfo {author} {\bibfnamefont {Davide}\ \bibnamefont
  {Gerosa}}\ and\ \bibinfo {author} {\bibfnamefont {Maya}\ \bibnamefont
  {Fishbach}},\ }\bibfield  {title} {\enquote {\bibinfo {title} {{Hierarchical
  mergers of stellar-mass black holes and their gravitational-wave
  signatures}},}\ }\href {\doibase 10.1038/s41550-021-01398-w} {\bibfield
  {journal} {\bibinfo  {journal} {Nature Astron.}\ }\textbf {\bibinfo {volume}
  {5}},\ \bibinfo {pages} {8} (\bibinfo {year} {2021})},\ \Eprint
  {http://arxiv.org/abs/2105.03439} {arXiv:2105.03439 [astro-ph.HE]}
  \BibitemShut {NoStop}%
\bibitem [{\citenamefont {Bogdanovic}\ \emph {et~al.}(2021)\citenamefont
  {Bogdanovic}, \citenamefont {Miller},\ and\ \citenamefont
  {Blecha}}]{Bogdanovic:2021aav}%
  \BibitemOpen
  \bibfield  {author} {\bibinfo {author} {\bibfnamefont {Tamara}\ \bibnamefont
  {Bogdanovic}}, \bibinfo {author} {\bibfnamefont {M.~Coleman}\ \bibnamefont
  {Miller}}, \ and\ \bibinfo {author} {\bibfnamefont {Laura}\ \bibnamefont
  {Blecha}},\ }\bibfield  {title} {\enquote {\bibinfo {title} {{Electromagnetic
  Counterparts to Massive Black Hole Mergers}},}\ }\href@noop {} {\  (\bibinfo
  {year} {2021})},\ \Eprint {http://arxiv.org/abs/2109.03262} {arXiv:2109.03262
  [astro-ph.HE]} \BibitemShut {NoStop}%
\bibitem [{\citenamefont {Varma}\ \emph {et~al.}(2020)\citenamefont {Varma},
  \citenamefont {Isi},\ and\ \citenamefont {Biscoveanu}}]{Varma:2020nbm}%
  \BibitemOpen
  \bibfield  {author} {\bibinfo {author} {\bibfnamefont {Vijay}\ \bibnamefont
  {Varma}}, \bibinfo {author} {\bibfnamefont {Maximiliano}\ \bibnamefont
  {Isi}}, \ and\ \bibinfo {author} {\bibfnamefont {Sylvia}\ \bibnamefont
  {Biscoveanu}},\ }\bibfield  {title} {\enquote {\bibinfo {title} {{Extracting
  the Gravitational Recoil from Black Hole Merger Signals}},}\ }\href {\doibase
  10.1103/PhysRevLett.124.101104} {\bibfield  {journal} {\bibinfo  {journal}
  {Phys. Rev. Lett.}\ }\textbf {\bibinfo {volume} {124}},\ \bibinfo {pages}
  {101104} (\bibinfo {year} {2020})},\ \Eprint
  {http://arxiv.org/abs/2002.00296} {arXiv:2002.00296 [gr-qc]} \BibitemShut
  {NoStop}%
\bibitem [{\citenamefont {Abbott}\ \emph
  {et~al.}(2020{\natexlab{b}})\citenamefont {Abbott} \emph
  {et~al.}}]{Abbott:2020mjq}%
  \BibitemOpen
  \bibfield  {author} {\bibinfo {author} {\bibfnamefont {R.}~\bibnamefont
  {Abbott}} \emph {et~al.} (\bibinfo {collaboration} {LIGO Scientific,
  Virgo}),\ }\bibfield  {title} {\enquote {\bibinfo {title} {{Properties and
  Astrophysical Implications of the 150 M$_\odot$ Binary Black Hole Merger
  GW190521}},}\ }\href {\doibase 10.3847/2041-8213/aba493} {\bibfield
  {journal} {\bibinfo  {journal} {Astrophys. J.}\ }\textbf {\bibinfo {volume}
  {900}},\ \bibinfo {pages} {L13} (\bibinfo {year} {2020}{\natexlab{b}})},\
  \Eprint {http://arxiv.org/abs/2009.01190} {arXiv:2009.01190 [astro-ph.HE]}
  \BibitemShut {NoStop}%
\bibitem [{\citenamefont {Gerosa}\ and\ \citenamefont
  {Moore}(2016)}]{Gerosa:2016vip}%
  \BibitemOpen
  \bibfield  {author} {\bibinfo {author} {\bibfnamefont {Davide}\ \bibnamefont
  {Gerosa}}\ and\ \bibinfo {author} {\bibfnamefont {Christopher~J.}\
  \bibnamefont {Moore}},\ }\bibfield  {title} {\enquote {\bibinfo {title}
  {{Black hole kicks as new gravitational wave observables}},}\ }\href
  {\doibase 10.1103/PhysRevLett.117.011101} {\bibfield  {journal} {\bibinfo
  {journal} {Phys. Rev. Lett.}\ }\textbf {\bibinfo {volume} {117}},\ \bibinfo
  {pages} {011101} (\bibinfo {year} {2016})},\ \Eprint
  {http://arxiv.org/abs/1606.04226} {arXiv:1606.04226 [gr-qc]} \BibitemShut
  {NoStop}%
\bibitem [{\citenamefont {Calderón~Bustillo}\ \emph
  {et~al.}(2018)\citenamefont {Calderón~Bustillo}, \citenamefont {Clark},
  \citenamefont {Laguna},\ and\ \citenamefont
  {Shoemaker}}]{CalderonBustillo:2018zuq}%
  \BibitemOpen
  \bibfield  {author} {\bibinfo {author} {\bibfnamefont {Juan}\ \bibnamefont
  {Calderón~Bustillo}}, \bibinfo {author} {\bibfnamefont {James~A.}\
  \bibnamefont {Clark}}, \bibinfo {author} {\bibfnamefont {Pablo}\ \bibnamefont
  {Laguna}}, \ and\ \bibinfo {author} {\bibfnamefont {Deirdre}\ \bibnamefont
  {Shoemaker}},\ }\bibfield  {title} {\enquote {\bibinfo {title} {{Tracking
  black hole kicks from gravitational wave observations}},}\ }\href {\doibase
  10.1103/PhysRevLett.121.191102} {\bibfield  {journal} {\bibinfo  {journal}
  {Phys. Rev. Lett.}\ }\textbf {\bibinfo {volume} {121}},\ \bibinfo {pages}
  {191102} (\bibinfo {year} {2018})},\ \Eprint
  {http://arxiv.org/abs/1806.11160} {arXiv:1806.11160 [gr-qc]} \BibitemShut
  {NoStop}%
\bibitem [{\citenamefont {Lousto}\ and\ \citenamefont
  {Healy}(2019)}]{Lousto:2019lyf}%
  \BibitemOpen
  \bibfield  {author} {\bibinfo {author} {\bibfnamefont {Carlos~O.}\
  \bibnamefont {Lousto}}\ and\ \bibinfo {author} {\bibfnamefont {James}\
  \bibnamefont {Healy}},\ }\bibfield  {title} {\enquote {\bibinfo {title}
  {{Kicking gravitational wave detectors with recoiling black holes}},}\ }\href
  {\doibase 10.1103/PhysRevD.100.104039} {\bibfield  {journal} {\bibinfo
  {journal} {Phys. Rev.}\ }\textbf {\bibinfo {volume} {D100}},\ \bibinfo
  {pages} {104039} (\bibinfo {year} {2019})},\ \Eprint
  {http://arxiv.org/abs/1908.04382} {arXiv:1908.04382 [gr-qc]} \BibitemShut
  {NoStop}%
\bibitem [{\citenamefont {Healy}\ \emph {et~al.}(2020)\citenamefont {Healy},
  \citenamefont {Lousto}, \citenamefont {Lange},\ and\ \citenamefont
  {O'Shaughnessy}}]{Healy:2020jjs}%
  \BibitemOpen
  \bibfield  {author} {\bibinfo {author} {\bibfnamefont {James}\ \bibnamefont
  {Healy}}, \bibinfo {author} {\bibfnamefont {Carlos~O.}\ \bibnamefont
  {Lousto}}, \bibinfo {author} {\bibfnamefont {Jacob}\ \bibnamefont {Lange}}, \
  and\ \bibinfo {author} {\bibfnamefont {Richard}\ \bibnamefont
  {O'Shaughnessy}},\ }\bibfield  {title} {\enquote {\bibinfo {title}
  {{Application of the third RIT binary black hole simulations catalog to
  parameter estimation of gravitational waves signals from the LIGO-Virgo O1/O2
  observational runs}},}\ }\href {\doibase 10.1103/PhysRevD.102.124053}
  {\bibfield  {journal} {\bibinfo  {journal} {Phys. Rev. D}\ }\textbf {\bibinfo
  {volume} {102}},\ \bibinfo {pages} {124053} (\bibinfo {year} {2020})},\
  \Eprint {http://arxiv.org/abs/2010.00108} {arXiv:2010.00108 [gr-qc]}
  \BibitemShut {NoStop}%
\bibitem [{\citenamefont {Mahapatra}\ \emph {et~al.}(2021)\citenamefont
  {Mahapatra}, \citenamefont {Gupta}, \citenamefont {Favata}, \citenamefont
  {Arun},\ and\ \citenamefont {Sathyaprakash}}]{Mahapatra:2021hme}%
  \BibitemOpen
  \bibfield  {author} {\bibinfo {author} {\bibfnamefont {Parthapratim}\
  \bibnamefont {Mahapatra}}, \bibinfo {author} {\bibfnamefont {Anuradha}\
  \bibnamefont {Gupta}}, \bibinfo {author} {\bibfnamefont {Marc}\ \bibnamefont
  {Favata}}, \bibinfo {author} {\bibfnamefont {K.~G.}\ \bibnamefont {Arun}}, \
  and\ \bibinfo {author} {\bibfnamefont {B.~S.}\ \bibnamefont
  {Sathyaprakash}},\ }\bibfield  {title} {\enquote {\bibinfo {title} {{Remnant
  Black Hole Kicks and Implications for Hierarchical Mergers}},}\ }\href
  {\doibase 10.3847/2041-8213/ac20db} {\bibfield  {journal} {\bibinfo
  {journal} {Astrophys. J. Lett.}\ }\textbf {\bibinfo {volume} {918}},\
  \bibinfo {pages} {L31} (\bibinfo {year} {2021})},\ \Eprint
  {http://arxiv.org/abs/2106.07179} {arXiv:2106.07179 [astro-ph.HE]}
  \BibitemShut {NoStop}%
\bibitem [{\citenamefont {Abbott}\ \emph {et~al.}(2019)\citenamefont {Abbott}
  \emph {et~al.}}]{LIGOScientific:2018mvr}%
  \BibitemOpen
  \bibfield  {author} {\bibinfo {author} {\bibfnamefont {B.~P.}\ \bibnamefont
  {Abbott}} \emph {et~al.} (\bibinfo {collaboration} {LIGO Scientific,
  Virgo}),\ }\bibfield  {title} {\enquote {\bibinfo {title} {{GWTC-1: A
  Gravitational-Wave Transient Catalog of Compact Binary Mergers Observed by
  LIGO and Virgo during the First and Second Observing Runs}},}\ }\href
  {\doibase 10.1103/PhysRevX.9.031040} {\bibfield  {journal} {\bibinfo
  {journal} {Phys. Rev.}\ }\textbf {\bibinfo {volume} {X9}},\ \bibinfo {pages}
  {031040} (\bibinfo {year} {2019})},\ \Eprint
  {http://arxiv.org/abs/1811.12907} {arXiv:1811.12907 [astro-ph.HE]}
  \BibitemShut {NoStop}%
\bibitem [{\citenamefont {Abbott}\ \emph
  {et~al.}(2021{\natexlab{c}})\citenamefont {Abbott} \emph
  {et~al.}}]{Abbott:2020gyp}%
  \BibitemOpen
  \bibfield  {author} {\bibinfo {author} {\bibfnamefont {R.}~\bibnamefont
  {Abbott}} \emph {et~al.} (\bibinfo {collaboration} {LIGO Scientific,
  Virgo}),\ }\bibfield  {title} {\enquote {\bibinfo {title} {{Population
  Properties of Compact Objects from the Second LIGO-Virgo Gravitational-Wave
  Transient Catalog}},}\ }\href {\doibase 10.3847/2041-8213/abe949} {\bibfield
  {journal} {\bibinfo  {journal} {Astrophys. J. Lett.}\ }\textbf {\bibinfo
  {volume} {913}},\ \bibinfo {pages} {L7} (\bibinfo {year}
  {2021}{\natexlab{c}})},\ \Eprint {http://arxiv.org/abs/2010.14533}
  {arXiv:2010.14533 [astro-ph.HE]} \BibitemShut {NoStop}%
\bibitem [{\citenamefont {Hannam}\ \emph {et~al.}(2021)\citenamefont {Hannam},
  \citenamefont {Hoy}, \citenamefont {Thompson}, \citenamefont {Fairhurst},
  \citenamefont {Raymond},\ and\ \citenamefont {LIGO}}]{Hannam:2021pit}%
  \BibitemOpen
  \bibfield  {author} {\bibinfo {author} {\bibfnamefont {Mark}\ \bibnamefont
  {Hannam}}, \bibinfo {author} {\bibfnamefont {Charlie}\ \bibnamefont {Hoy}},
  \bibinfo {author} {\bibfnamefont {Jonathan~E.}\ \bibnamefont {Thompson}},
  \bibinfo {author} {\bibfnamefont {Stephen}\ \bibnamefont {Fairhurst}},
  \bibinfo {author} {\bibfnamefont {Vivien}\ \bibnamefont {Raymond}}, \ and\
  \bibinfo {author} {\bibfnamefont {members of~the}\ \bibnamefont {LIGO}}
  (\bibinfo {collaboration} {VIRGO}),\ }\bibfield  {title} {\enquote {\bibinfo
  {title} {{Measurement of general-relativistic precession in a black-hole
  binary}},}\ }\href@noop {} {\  (\bibinfo {year} {2021})},\ \Eprint
  {http://arxiv.org/abs/2112.11300} {arXiv:2112.11300 [gr-qc]} \BibitemShut
  {NoStop}%
\bibitem [{\citenamefont {Varma}\ \emph
  {et~al.}(2021{\natexlab{a}})\citenamefont {Varma}, \citenamefont
  {Biscoveanu}, \citenamefont {Isi}, \citenamefont {Farr},\ and\ \citenamefont
  {Vitale}}]{Varma:2021xbh}%
  \BibitemOpen
  \bibfield  {author} {\bibinfo {author} {\bibfnamefont {Vijay}\ \bibnamefont
  {Varma}}, \bibinfo {author} {\bibfnamefont {Sylvia}\ \bibnamefont
  {Biscoveanu}}, \bibinfo {author} {\bibfnamefont {Maximiliano}\ \bibnamefont
  {Isi}}, \bibinfo {author} {\bibfnamefont {Will~M.}\ \bibnamefont {Farr}}, \
  and\ \bibinfo {author} {\bibfnamefont {Salvatore}\ \bibnamefont {Vitale}},\
  }\bibfield  {title} {\enquote {\bibinfo {title} {{Hints of spin-orbit
  resonances in the binary black hole population}},}\ }\href@noop {} {\
  (\bibinfo {year} {2021}{\natexlab{a}})},\ \Eprint
  {http://arxiv.org/abs/2107.09693} {arXiv:2107.09693 [astro-ph.HE]}
  \BibitemShut {NoStop}%
\bibitem [{\citenamefont {Doctor}\ \emph {et~al.}(2021)\citenamefont {Doctor},
  \citenamefont {Farr},\ and\ \citenamefont {Holz}}]{Doctor:2021qfn}%
  \BibitemOpen
  \bibfield  {author} {\bibinfo {author} {\bibfnamefont {Zoheyr}\ \bibnamefont
  {Doctor}}, \bibinfo {author} {\bibfnamefont {Ben}\ \bibnamefont {Farr}}, \
  and\ \bibinfo {author} {\bibfnamefont {Daniel~E.}\ \bibnamefont {Holz}},\
  }\bibfield  {title} {\enquote {\bibinfo {title} {{Black Hole Leftovers: The
  Remnant Population from Binary Black Hole Mergers}},}\ }\href {\doibase
  10.3847/2041-8213/ac0334} {\bibfield  {journal} {\bibinfo  {journal}
  {Astrophys. J. Lett.}\ }\textbf {\bibinfo {volume} {914}},\ \bibinfo {pages}
  {L18} (\bibinfo {year} {2021})},\ \Eprint {http://arxiv.org/abs/2103.04001}
  {arXiv:2103.04001 [astro-ph.HE]} \BibitemShut {NoStop}%
\bibitem [{\citenamefont {Islam}\ \emph {et~al.}(2022)\citenamefont {Islam},
  \citenamefont {Shaik}, \citenamefont {Haster}, \citenamefont {Varma},
  \citenamefont {Field}, \citenamefont {Lange}, \citenamefont
  {O’Shaughnessy}, \citenamefont {Smith},\ and\ \citenamefont
  {Vajpeyi}}]{Islam:2022_inprep}%
  \BibitemOpen
  \bibfield  {author} {\bibinfo {author} {\bibfnamefont {Tousif}\ \bibnamefont
  {Islam}}, \bibinfo {author} {\bibfnamefont {Feroz}\ \bibnamefont {Shaik}},
  \bibinfo {author} {\bibfnamefont {Carl-Johan}\ \bibnamefont {Haster}},
  \bibinfo {author} {\bibfnamefont {Vijay}\ \bibnamefont {Varma}}, \bibinfo
  {author} {\bibfnamefont {Scott}\ \bibnamefont {Field}}, \bibinfo {author}
  {\bibfnamefont {Jacob}\ \bibnamefont {Lange}}, \bibinfo {author}
  {\bibfnamefont {Richard}\ \bibnamefont {O’Shaughnessy}}, \bibinfo {author}
  {\bibfnamefont {Rory}\ \bibnamefont {Smith}}, \ and\ \bibinfo {author}
  {\bibfnamefont {Avi}\ \bibnamefont {Vajpeyi}},\ }\bibfield  {title} {\enquote
  {\bibinfo {title} {{Re-analysis of GWTC-3 events with Numerical Relativity
  Surrogate models}},}\ }\href@noop {} {\  (\bibinfo {year} {2022})},\ \bibinfo
  {note} {in preparation}\BibitemShut {NoStop}%
\bibitem [{\citenamefont {Abbott}\ \emph
  {et~al.}(2020{\natexlab{c}})\citenamefont {Abbott} \emph
  {et~al.}}]{Abbott:2020khf}%
  \BibitemOpen
  \bibfield  {author} {\bibinfo {author} {\bibfnamefont {R.}~\bibnamefont
  {Abbott}} \emph {et~al.} (\bibinfo {collaboration} {LIGO Scientific,
  Virgo}),\ }\bibfield  {title} {\enquote {\bibinfo {title} {{GW190814:
  Gravitational Waves from the Coalescence of a 23 Solar Mass Black Hole with a
  2.6 Solar Mass Compact Object}},}\ }\href {\doibase 10.3847/2041-8213/ab960f}
  {\bibfield  {journal} {\bibinfo  {journal} {Astrophys. J. Lett.}\ }\textbf
  {\bibinfo {volume} {896}},\ \bibinfo {pages} {L44} (\bibinfo {year}
  {2020}{\natexlab{c}})},\ \Eprint {http://arxiv.org/abs/2006.12611}
  {arXiv:2006.12611 [astro-ph.HE]} \BibitemShut {NoStop}%
\bibitem [{\citenamefont {{Thrane}}\ and\ \citenamefont
  {{Talbot}}(2019)}]{Thrane:2019pe}%
  \BibitemOpen
  \bibfield  {author} {\bibinfo {author} {\bibfnamefont {Eric}\ \bibnamefont
  {{Thrane}}}\ and\ \bibinfo {author} {\bibfnamefont {Colm}\ \bibnamefont
  {{Talbot}}},\ }\bibfield  {title} {\enquote {\bibinfo {title} {{An
  introduction to Bayesian inference in gravitational-wave astronomy: Parameter
  estimation, model selection, and hierarchical models}},}\ }\href {\doibase
  10.1017/pasa.2019.2} {\bibfield  {journal} {\bibinfo  {journal} {Publications
  of the Astronomical Society of Australia}\ }\textbf {\bibinfo {volume}
  {36}},\ \bibinfo {eid} {e010} (\bibinfo {year} {2019})},\ \Eprint
  {http://arxiv.org/abs/1809.02293} {arXiv:1809.02293 [astro-ph.IM]}
  \BibitemShut {NoStop}%
\bibitem [{\citenamefont {Smith}\ \emph {et~al.}(2020)\citenamefont {Smith},
  \citenamefont {Ashton}, \citenamefont {Vajpeyi},\ and\ \citenamefont
  {Talbot}}]{Smith:2019ucc}%
  \BibitemOpen
  \bibfield  {author} {\bibinfo {author} {\bibfnamefont {Rory~J.E.}\
  \bibnamefont {Smith}}, \bibinfo {author} {\bibfnamefont {Gregory}\
  \bibnamefont {Ashton}}, \bibinfo {author} {\bibfnamefont {Avi}\ \bibnamefont
  {Vajpeyi}}, \ and\ \bibinfo {author} {\bibfnamefont {Colm}\ \bibnamefont
  {Talbot}},\ }\bibfield  {title} {\enquote {\bibinfo {title} {{Massively
  parallel Bayesian inference for transient gravitational-wave astronomy}},}\
  }\href {\doibase 10.1093/mnras/staa2483} {\bibfield  {journal} {\bibinfo
  {journal} {Mon. Not. Roy. Astron. Soc.}\ }\textbf {\bibinfo {volume} {498}},\
  \bibinfo {pages} {4492--4502} (\bibinfo {year} {2020})},\ \Eprint
  {http://arxiv.org/abs/1909.11873} {arXiv:1909.11873 [gr-qc]} \BibitemShut
  {NoStop}%
\bibitem [{\citenamefont {{Speagle}}(2020)}]{Speagle:2019dynesty}%
  \BibitemOpen
  \bibfield  {author} {\bibinfo {author} {\bibfnamefont {Joshua~S.}\
  \bibnamefont {{Speagle}}},\ }\bibfield  {title} {\enquote {\bibinfo {title}
  {{DYNESTY: a dynamic nested sampling package for estimating Bayesian
  posteriors and evidences}},}\ }\href {\doibase 10.1093/mnras/staa278}
  {\bibfield  {journal} {\bibinfo  {journal} {Monthly Notices of the Royal
  Astronomical Society}\ }\textbf {\bibinfo {volume} {493}},\ \bibinfo {pages}
  {3132--3158} (\bibinfo {year} {2020})},\ \Eprint
  {http://arxiv.org/abs/1904.02180} {arXiv:1904.02180 [astro-ph.IM]}
  \BibitemShut {NoStop}%
\bibitem [{\citenamefont {Varma}\ \emph
  {et~al.}(2019{\natexlab{a}})\citenamefont {Varma}, \citenamefont {Field},
  \citenamefont {Scheel}, \citenamefont {Blackman}, \citenamefont {Gerosa},
  \citenamefont {Stein}, \citenamefont {Kidder},\ and\ \citenamefont
  {Pfeiffer}}]{Varma:2019csw}%
  \BibitemOpen
  \bibfield  {author} {\bibinfo {author} {\bibfnamefont {Vijay}\ \bibnamefont
  {Varma}}, \bibinfo {author} {\bibfnamefont {Scott~E.}\ \bibnamefont {Field}},
  \bibinfo {author} {\bibfnamefont {Mark~A.}\ \bibnamefont {Scheel}}, \bibinfo
  {author} {\bibfnamefont {Jonathan}\ \bibnamefont {Blackman}}, \bibinfo
  {author} {\bibfnamefont {Davide}\ \bibnamefont {Gerosa}}, \bibinfo {author}
  {\bibfnamefont {Leo~C.}\ \bibnamefont {Stein}}, \bibinfo {author}
  {\bibfnamefont {Lawrence~E.}\ \bibnamefont {Kidder}}, \ and\ \bibinfo
  {author} {\bibfnamefont {Harald~P.}\ \bibnamefont {Pfeiffer}},\ }\bibfield
  {title} {\enquote {\bibinfo {title} {{Surrogate models for precessing binary
  black hole simulations with unequal masses}},}\ }\href {\doibase
  10.1103/PhysRevResearch.1.033015} {\bibfield  {journal} {\bibinfo  {journal}
  {Phys. Rev. Research.}\ }\textbf {\bibinfo {volume} {1}},\ \bibinfo {pages}
  {033015} (\bibinfo {year} {2019}{\natexlab{a}})},\ \Eprint
  {http://arxiv.org/abs/1905.09300} {arXiv:1905.09300 [gr-qc]} \BibitemShut
  {NoStop}%
\bibitem [{\citenamefont {Varma}\ \emph
  {et~al.}(2019{\natexlab{b}})\citenamefont {Varma}, \citenamefont {Gerosa},
  \citenamefont {Stein}, \citenamefont {Hébert},\ and\ \citenamefont
  {Zhang}}]{Varma:2018aht}%
  \BibitemOpen
  \bibfield  {author} {\bibinfo {author} {\bibfnamefont {Vijay}\ \bibnamefont
  {Varma}}, \bibinfo {author} {\bibfnamefont {Davide}\ \bibnamefont {Gerosa}},
  \bibinfo {author} {\bibfnamefont {Leo~C.}\ \bibnamefont {Stein}}, \bibinfo
  {author} {\bibfnamefont {François}\ \bibnamefont {Hébert}}, \ and\ \bibinfo
  {author} {\bibfnamefont {Hao}\ \bibnamefont {Zhang}},\ }\bibfield  {title}
  {\enquote {\bibinfo {title} {{High-accuracy mass, spin, and recoil
  predictions of generic black-hole merger remnants}},}\ }\href {\doibase
  10.1103/PhysRevLett.122.011101} {\bibfield  {journal} {\bibinfo  {journal}
  {Phys. Rev. Lett.}\ }\textbf {\bibinfo {volume} {122}},\ \bibinfo {pages}
  {011101} (\bibinfo {year} {2019}{\natexlab{b}})},\ \Eprint
  {http://arxiv.org/abs/1809.09125} {arXiv:1809.09125 [gr-qc]} \BibitemShut
  {NoStop}%
\bibitem [{\citenamefont {Pratten}\ \emph {et~al.}(2021)\citenamefont {Pratten}
  \emph {et~al.}}]{Pratten:2020ceb}%
  \BibitemOpen
  \bibfield  {author} {\bibinfo {author} {\bibfnamefont {Geraint}\ \bibnamefont
  {Pratten}} \emph {et~al.},\ }\bibfield  {title} {\enquote {\bibinfo {title}
  {{Computationally efficient models for the dominant and subdominant harmonic
  modes of precessing binary black holes}},}\ }\href {\doibase
  10.1103/PhysRevD.103.104056} {\bibfield  {journal} {\bibinfo  {journal}
  {Phys. Rev. D}\ }\textbf {\bibinfo {volume} {103}},\ \bibinfo {pages}
  {104056} (\bibinfo {year} {2021})},\ \Eprint
  {http://arxiv.org/abs/2004.06503} {arXiv:2004.06503 [gr-qc]} \BibitemShut
  {NoStop}%
\bibitem [{\citenamefont {Ossokine}\ \emph {et~al.}(2020)\citenamefont
  {Ossokine} \emph {et~al.}}]{Ossokine:2020kjp}%
  \BibitemOpen
  \bibfield  {author} {\bibinfo {author} {\bibfnamefont {Serguei}\ \bibnamefont
  {Ossokine}} \emph {et~al.},\ }\bibfield  {title} {\enquote {\bibinfo {title}
  {{Multipolar Effective-One-Body Waveforms for Precessing Binary Black Holes:
  Construction and Validation}},}\ }\href {\doibase
  10.1103/PhysRevD.102.044055} {\bibfield  {journal} {\bibinfo  {journal}
  {Phys. Rev. D}\ }\textbf {\bibinfo {volume} {102}},\ \bibinfo {pages}
  {044055} (\bibinfo {year} {2020})},\ \Eprint
  {http://arxiv.org/abs/2004.09442} {arXiv:2004.09442 [gr-qc]} \BibitemShut
  {NoStop}%
\bibitem [{\citenamefont {Hofmann}\ \emph {et~al.}(2016)\citenamefont
  {Hofmann}, \citenamefont {Barausse},\ and\ \citenamefont
  {Rezzolla}}]{Hofmann:2016yih}%
  \BibitemOpen
  \bibfield  {author} {\bibinfo {author} {\bibfnamefont {Fabian}\ \bibnamefont
  {Hofmann}}, \bibinfo {author} {\bibfnamefont {Enrico}\ \bibnamefont
  {Barausse}}, \ and\ \bibinfo {author} {\bibfnamefont {Luciano}\ \bibnamefont
  {Rezzolla}},\ }\bibfield  {title} {\enquote {\bibinfo {title} {{The final
  spin from binary black holes in quasi-circular orbits}},}\ }\href {\doibase
  10.3847/2041-8205/825/2/L19} {\bibfield  {journal} {\bibinfo  {journal}
  {Astrophys. J.}\ }\textbf {\bibinfo {volume} {825}},\ \bibinfo {pages} {L19}
  (\bibinfo {year} {2016})},\ \Eprint {http://arxiv.org/abs/1605.01938}
  {arXiv:1605.01938 [gr-qc]} \BibitemShut {NoStop}%
\bibitem [{\citenamefont {Barausse}\ \emph {et~al.}(2012)\citenamefont
  {Barausse}, \citenamefont {Morozova},\ and\ \citenamefont
  {Rezzolla}}]{Barausse:2012qz}%
  \BibitemOpen
  \bibfield  {author} {\bibinfo {author} {\bibfnamefont {Enrico}\ \bibnamefont
  {Barausse}}, \bibinfo {author} {\bibfnamefont {Viktoriya}\ \bibnamefont
  {Morozova}}, \ and\ \bibinfo {author} {\bibfnamefont {Luciano}\ \bibnamefont
  {Rezzolla}},\ }\bibfield  {title} {\enquote {\bibinfo {title} {{On the mass
  radiated by coalescing black-hole binaries}},}\ }\href {\doibase
  10.1088/0004-637X/758/1/63} {\bibfield  {journal} {\bibinfo  {journal}
  {Astrophys. J.}\ }\textbf {\bibinfo {volume} {758}},\ \bibinfo {pages} {63}
  (\bibinfo {year} {2012})},\ \bibinfo {note} {[Erratum: Astrophys.
  J.786,76(2014)]},\ \Eprint {http://arxiv.org/abs/1206.3803} {arXiv:1206.3803
  [gr-qc]} \BibitemShut {NoStop}%
\bibitem [{\citenamefont {Jiménez-Forteza}\ \emph {et~al.}(2017)\citenamefont
  {Jiménez-Forteza}, \citenamefont {Keitel}, \citenamefont {Husa},
  \citenamefont {Hannam}, \citenamefont {Khan},\ and\ \citenamefont
  {Pürrer}}]{Jimenez-Forteza:2016oae}%
  \BibitemOpen
  \bibfield  {author} {\bibinfo {author} {\bibfnamefont {Xisco}\ \bibnamefont
  {Jiménez-Forteza}}, \bibinfo {author} {\bibfnamefont {David}\ \bibnamefont
  {Keitel}}, \bibinfo {author} {\bibfnamefont {Sascha}\ \bibnamefont {Husa}},
  \bibinfo {author} {\bibfnamefont {Mark}\ \bibnamefont {Hannam}}, \bibinfo
  {author} {\bibfnamefont {Sebastian}\ \bibnamefont {Khan}}, \ and\ \bibinfo
  {author} {\bibfnamefont {Michael}\ \bibnamefont {Pürrer}},\ }\bibfield
  {title} {\enquote {\bibinfo {title} {{Hierarchical data-driven approach to
  fitting numerical relativity data for nonprecessing binary black holes with
  an application to final spin and radiated energy}},}\ }\href {\doibase
  10.1103/PhysRevD.95.064024} {\bibfield  {journal} {\bibinfo  {journal} {Phys.
  Rev.}\ }\textbf {\bibinfo {volume} {D95}},\ \bibinfo {pages} {064024}
  (\bibinfo {year} {2017})},\ \Eprint {http://arxiv.org/abs/1611.00332}
  {arXiv:1611.00332 [gr-qc]} \BibitemShut {NoStop}%
\bibitem [{\citenamefont {Lousto}\ \emph {et~al.}(2012)\citenamefont {Lousto},
  \citenamefont {Zlochower}, \citenamefont {Dotti},\ and\ \citenamefont
  {Volonteri}}]{Lousto:2012su}%
  \BibitemOpen
  \bibfield  {author} {\bibinfo {author} {\bibfnamefont {Carlos~O.}\
  \bibnamefont {Lousto}}, \bibinfo {author} {\bibfnamefont {Yosef}\
  \bibnamefont {Zlochower}}, \bibinfo {author} {\bibfnamefont {Massimo}\
  \bibnamefont {Dotti}}, \ and\ \bibinfo {author} {\bibfnamefont {Marta}\
  \bibnamefont {Volonteri}},\ }\bibfield  {title} {\enquote {\bibinfo {title}
  {{Gravitational Recoil From Accretion-Aligned Black-Hole Binaries}},}\ }\href
  {\doibase 10.1103/PhysRevD.85.084015} {\bibfield  {journal} {\bibinfo
  {journal} {Phys. Rev.}\ }\textbf {\bibinfo {volume} {D85}},\ \bibinfo {pages}
  {084015} (\bibinfo {year} {2012})},\ \Eprint {http://arxiv.org/abs/1201.1923}
  {arXiv:1201.1923 [gr-qc]} \BibitemShut {NoStop}%
\bibitem [{\citenamefont {Varma}\ \emph
  {et~al.}(2019{\natexlab{c}})\citenamefont {Varma}, \citenamefont {Field},
  \citenamefont {Scheel}, \citenamefont {Blackman}, \citenamefont {Kidder},\
  and\ \citenamefont {Pfeiffer}}]{Varma:2018mmi}%
  \BibitemOpen
  \bibfield  {author} {\bibinfo {author} {\bibfnamefont {Vijay}\ \bibnamefont
  {Varma}}, \bibinfo {author} {\bibfnamefont {Scott~E.}\ \bibnamefont {Field}},
  \bibinfo {author} {\bibfnamefont {Mark~A.}\ \bibnamefont {Scheel}}, \bibinfo
  {author} {\bibfnamefont {Jonathan}\ \bibnamefont {Blackman}}, \bibinfo
  {author} {\bibfnamefont {Lawrence~E.}\ \bibnamefont {Kidder}}, \ and\
  \bibinfo {author} {\bibfnamefont {Harald~P.}\ \bibnamefont {Pfeiffer}},\
  }\bibfield  {title} {\enquote {\bibinfo {title} {{Surrogate model of
  hybridized numerical relativity binary black hole waveforms}},}\ }\href
  {\doibase 10.1103/PhysRevD.99.064045} {\bibfield  {journal} {\bibinfo
  {journal} {Phys. Rev.}\ }\textbf {\bibinfo {volume} {D99}},\ \bibinfo {pages}
  {064045} (\bibinfo {year} {2019}{\natexlab{c}})},\ \Eprint
  {http://arxiv.org/abs/1812.07865} {arXiv:1812.07865 [gr-qc]} \BibitemShut
  {NoStop}%
\bibitem [{\citenamefont {Blackman}\ \emph {et~al.}(2017)\citenamefont
  {Blackman}, \citenamefont {Field}, \citenamefont {Scheel}, \citenamefont
  {Galley}, \citenamefont {Ott}, \citenamefont {Boyle}, \citenamefont {Kidder},
  \citenamefont {Pfeiffer},\ and\ \citenamefont
  {Szilágyi}}]{Blackman:2017pcm}%
  \BibitemOpen
  \bibfield  {author} {\bibinfo {author} {\bibfnamefont {Jonathan}\
  \bibnamefont {Blackman}}, \bibinfo {author} {\bibfnamefont {Scott~E.}\
  \bibnamefont {Field}}, \bibinfo {author} {\bibfnamefont {Mark~A.}\
  \bibnamefont {Scheel}}, \bibinfo {author} {\bibfnamefont {Chad~R.}\
  \bibnamefont {Galley}}, \bibinfo {author} {\bibfnamefont {Christian~D.}\
  \bibnamefont {Ott}}, \bibinfo {author} {\bibfnamefont {Michael}\ \bibnamefont
  {Boyle}}, \bibinfo {author} {\bibfnamefont {Lawrence~E.}\ \bibnamefont
  {Kidder}}, \bibinfo {author} {\bibfnamefont {Harald~P.}\ \bibnamefont
  {Pfeiffer}}, \ and\ \bibinfo {author} {\bibfnamefont {Béla}\ \bibnamefont
  {Szilágyi}},\ }\bibfield  {title} {\enquote {\bibinfo {title} {{Numerical
  relativity waveform surrogate model for generically precessing binary black
  hole mergers}},}\ }\href {\doibase 10.1103/PhysRevD.96.024058} {\bibfield
  {journal} {\bibinfo  {journal} {Phys. Rev.}\ }\textbf {\bibinfo {volume}
  {D96}},\ \bibinfo {pages} {024058} (\bibinfo {year} {2017})},\ \Eprint
  {http://arxiv.org/abs/1705.07089} {arXiv:1705.07089 [gr-qc]} \BibitemShut
  {NoStop}%
\bibitem [{kic()}]{kickpapersupplement}%
  \BibitemOpen
  \href@noop {} {}\bibinfo {howpublished} {See Supplemental Material
  \hyperlink{page.9}{here}, for tests of the surrogate models using high spin
  NR injections. This further includes Refs.~\cite{Boyle:2019kee,
  Beth:2021_inprep, Scheel:2014ina, aLIGODesignNoiseCurve, SXSCatalog,
  Chatziioannou:2018wqx}.}\BibitemShut {Stop}%
\bibitem [{\citenamefont {Romero-Shaw}\ \emph {et~al.}(2020)\citenamefont
  {Romero-Shaw} \emph {et~al.}}]{Romero-Shaw:2020owr}%
  \BibitemOpen
  \bibfield  {author} {\bibinfo {author} {\bibfnamefont {I.~M.}\ \bibnamefont
  {Romero-Shaw}} \emph {et~al.},\ }\bibfield  {title} {\enquote {\bibinfo
  {title} {{Bayesian inference for compact binary coalescences with bilby:
  validation and application to the first LIGO\textendash{}Virgo
  gravitational-wave transient catalogue}},}\ }\href {\doibase
  10.1093/mnras/staa2850} {\bibfield  {journal} {\bibinfo  {journal} {Mon. Not.
  Roy. Astron. Soc.}\ }\textbf {\bibinfo {volume} {499}},\ \bibinfo {pages}
  {3295--3319} (\bibinfo {year} {2020})},\ \Eprint
  {http://arxiv.org/abs/2006.00714} {arXiv:2006.00714 [astro-ph.IM]}
  \BibitemShut {NoStop}%
\bibitem [{\citenamefont {Varma}\ \emph
  {et~al.}(2021{\natexlab{b}})\citenamefont {Varma}, \citenamefont {Isi},
  \citenamefont {Biscoveanu}, \citenamefont {Farr},\ and\ \citenamefont
  {Vitale}}]{Varma:2021csh}%
  \BibitemOpen
  \bibfield  {author} {\bibinfo {author} {\bibfnamefont {Vijay}\ \bibnamefont
  {Varma}}, \bibinfo {author} {\bibfnamefont {Maximiliano}\ \bibnamefont
  {Isi}}, \bibinfo {author} {\bibfnamefont {Sylvia}\ \bibnamefont
  {Biscoveanu}}, \bibinfo {author} {\bibfnamefont {Will~M.}\ \bibnamefont
  {Farr}}, \ and\ \bibinfo {author} {\bibfnamefont {Salvatore}\ \bibnamefont
  {Vitale}},\ }\bibfield  {title} {\enquote {\bibinfo {title} {{Measuring
  binary black hole orbital-plane spin orientations}},}\ }\href@noop {} {\
  (\bibinfo {year} {2021}{\natexlab{b}})},\ \Eprint
  {http://arxiv.org/abs/2107.09692} {arXiv:2107.09692 [astro-ph.HE]}
  \BibitemShut {NoStop}%
\bibitem [{\citenamefont {Antonini}\ and\ \citenamefont
  {Rasio}(2016)}]{Antonini:2016gqe}%
  \BibitemOpen
  \bibfield  {author} {\bibinfo {author} {\bibfnamefont {Fabio}\ \bibnamefont
  {Antonini}}\ and\ \bibinfo {author} {\bibfnamefont {Frederic~A.}\
  \bibnamefont {Rasio}},\ }\bibfield  {title} {\enquote {\bibinfo {title}
  {{Merging black hole binaries in galactic nuclei: implications for
  advanced-LIGO detections}},}\ }\href {\doibase 10.3847/0004-637X/831/2/187}
  {\bibfield  {journal} {\bibinfo  {journal} {Astrophys. J.}\ }\textbf
  {\bibinfo {volume} {831}},\ \bibinfo {pages} {187} (\bibinfo {year}
  {2016})},\ \Eprint {http://arxiv.org/abs/1606.04889} {arXiv:1606.04889
  [astro-ph.HE]} \BibitemShut {NoStop}%
\bibitem [{\citenamefont {{Monari}}\ \emph {et~al.}(2018)\citenamefont
  {{Monari}}, \citenamefont {{Famaey}}, \citenamefont {{Carrillo}},
  \citenamefont {{Piffl}}, \citenamefont {{Steinmetz}}, \citenamefont {{Wyse}},
  \citenamefont {{Anders}}, \citenamefont {{Chiappini}},\ and\ \citenamefont
  {{Jan{\ss}en}}}]{Monari:2018esc}%
  \BibitemOpen
  \bibfield  {author} {\bibinfo {author} {\bibfnamefont {G.}~\bibnamefont
  {{Monari}}}, \bibinfo {author} {\bibfnamefont {B.}~\bibnamefont {{Famaey}}},
  \bibinfo {author} {\bibfnamefont {I.}~\bibnamefont {{Carrillo}}}, \bibinfo
  {author} {\bibfnamefont {T.}~\bibnamefont {{Piffl}}}, \bibinfo {author}
  {\bibfnamefont {M.}~\bibnamefont {{Steinmetz}}}, \bibinfo {author}
  {\bibfnamefont {R.~F.~G.}\ \bibnamefont {{Wyse}}}, \bibinfo {author}
  {\bibfnamefont {F.}~\bibnamefont {{Anders}}}, \bibinfo {author}
  {\bibfnamefont {C.}~\bibnamefont {{Chiappini}}}, \ and\ \bibinfo {author}
  {\bibfnamefont {K.}~\bibnamefont {{Jan{\ss}en}}},\ }\bibfield  {title}
  {\enquote {\bibinfo {title} {{The escape speed curve of the Galaxy obtained
  from Gaia DR2 implies a heavy Milky Way}},}\ }\href {\doibase
  10.1051/0004-6361/201833748} {\bibfield  {journal} {\bibinfo  {journal}
  {A\&A}\ }\textbf {\bibinfo {volume} {616}},\ \bibinfo {eid} {L9} (\bibinfo
  {year} {2018})},\ \Eprint {http://arxiv.org/abs/1807.04565} {arXiv:1807.04565
  [astro-ph.GA]} \BibitemShut {NoStop}%
\bibitem [{\citenamefont {Kullback}\ and\ \citenamefont
  {Leibler}(1951)}]{Kullback_Leibler_divergence}%
  \BibitemOpen
  \bibfield  {author} {\bibinfo {author} {\bibfnamefont {S.}~\bibnamefont
  {Kullback}}\ and\ \bibinfo {author} {\bibfnamefont {R.~A.}\ \bibnamefont
  {Leibler}},\ }\bibfield  {title} {\enquote {\bibinfo {title} {On information
  and sufficiency},}\ }\href {http://www.jstor.org/stable/2236703} {\bibfield
  {journal} {\bibinfo  {journal} {The Annals of Mathematical Statistics}\
  }\textbf {\bibinfo {volume} {22}},\ \bibinfo {pages} {79--86} (\bibinfo
  {year} {1951})}\BibitemShut {NoStop}%
\bibitem [{\citenamefont {Ma}\ \emph {et~al.}(2021)\citenamefont {Ma},
  \citenamefont {Giesler}, \citenamefont {Varma}, \citenamefont {Scheel},\ and\
  \citenamefont {Chen}}]{Ma:2021znq}%
  \BibitemOpen
  \bibfield  {author} {\bibinfo {author} {\bibfnamefont {Sizheng}\ \bibnamefont
  {Ma}}, \bibinfo {author} {\bibfnamefont {Matthew}\ \bibnamefont {Giesler}},
  \bibinfo {author} {\bibfnamefont {Vijay}\ \bibnamefont {Varma}}, \bibinfo
  {author} {\bibfnamefont {Mark~A.}\ \bibnamefont {Scheel}}, \ and\ \bibinfo
  {author} {\bibfnamefont {Yanbei}\ \bibnamefont {Chen}},\ }\bibfield  {title}
  {\enquote {\bibinfo {title} {{Universal features of gravitational waves
  emitted by superkick binary black hole systems}},}\ }\href {\doibase
  10.1103/PhysRevD.104.084003} {\bibfield  {journal} {\bibinfo  {journal}
  {Phys. Rev. D}\ }\textbf {\bibinfo {volume} {104}},\ \bibinfo {pages}
  {084003} (\bibinfo {year} {2021})},\ \Eprint
  {http://arxiv.org/abs/2107.04890} {arXiv:2107.04890 [gr-qc]} \BibitemShut
  {NoStop}%
\bibitem [{\citenamefont {Abbott}\ \emph
  {et~al.}(2021{\natexlab{d}})\citenamefont {Abbott} \emph
  {et~al.}}]{LIGOScientific:2021sio}%
  \BibitemOpen
  \bibfield  {author} {\bibinfo {author} {\bibfnamefont {R.}~\bibnamefont
  {Abbott}} \emph {et~al.} (\bibinfo {collaboration} {LIGO Scientific, VIRGO,
  KAGRA}),\ }\bibfield  {title} {\enquote {\bibinfo {title} {{Tests of General
  Relativity with GWTC-3}},}\ }\href@noop {} {\  (\bibinfo {year}
  {2021}{\natexlab{d}})},\ \Eprint {http://arxiv.org/abs/2112.06861}
  {arXiv:2112.06861 [gr-qc]} \BibitemShut {NoStop}%
\bibitem [{\citenamefont {Collaboration}\ and\ \citenamefont
  {Collaboration}()}]{GW_open_science_center}%
  \BibitemOpen
  \bibfield  {author} {\bibinfo {author} {\bibfnamefont {LIGO~Scientific}\
  \bibnamefont {Collaboration}}\ and\ \bibinfo {author} {\bibfnamefont {Virgo}\
  \bibnamefont {Collaboration}},\ }\bibfield  {title} {\enquote {\bibinfo
  {title} {{Gravitational Wave Open Science Center}},}\ }\href@noop {} {\
  }\bibinfo {note} {\url{https://www.gw-openscience.org}}\BibitemShut {NoStop}%
\bibitem [{\citenamefont {Boyle}\ \emph {et~al.}(2019)\citenamefont {Boyle}
  \emph {et~al.}}]{Boyle:2019kee}%
  \BibitemOpen
  \bibfield  {author} {\bibinfo {author} {\bibfnamefont {Michael}\ \bibnamefont
  {Boyle}} \emph {et~al.},\ }\bibfield  {title} {\enquote {\bibinfo {title}
  {{The SXS Collaboration catalog of binary black hole simulations}},}\ }\href
  {\doibase 10.1088/1361-6382/ab34e2} {\bibfield  {journal} {\bibinfo
  {journal} {Class. Quant. Grav.}\ }\textbf {\bibinfo {volume} {36}},\ \bibinfo
  {pages} {195006} (\bibinfo {year} {2019})},\ \Eprint
  {http://arxiv.org/abs/1904.04831} {arXiv:1904.04831 [gr-qc]} \BibitemShut
  {NoStop}%
\bibitem [{\citenamefont {Walker}\ \emph {et~al.}(2021)\citenamefont {Walker},
  \citenamefont {Varma},\ and\ \citenamefont {Lovelace}}]{Beth:2021_inprep}%
  \BibitemOpen
  \bibfield  {author} {\bibinfo {author} {\bibfnamefont {Marissa}\ \bibnamefont
  {Walker}}, \bibinfo {author} {\bibfnamefont {Vijay}\ \bibnamefont {Varma}}, \
  and\ \bibinfo {author} {\bibfnamefont {Geoffrey}\ \bibnamefont {Lovelace}},\
  }\bibfield  {title} {\enquote {\bibinfo {title} {{Extending numerical
  relativity surrogate models to near extremal spins}},}\ }\href@noop {} {\
  (\bibinfo {year} {2021})},\ \bibinfo {note} {in preparation}\BibitemShut
  {NoStop}%
\bibitem [{\citenamefont {Scheel}\ \emph {et~al.}(2015)\citenamefont {Scheel},
  \citenamefont {Giesler}, \citenamefont {Hemberger}, \citenamefont {Lovelace},
  \citenamefont {Kuper}, \citenamefont {Boyle}, \citenamefont {Szilágyi},\
  and\ \citenamefont {Kidder}}]{Scheel:2014ina}%
  \BibitemOpen
  \bibfield  {author} {\bibinfo {author} {\bibfnamefont {Mark~A.}\ \bibnamefont
  {Scheel}}, \bibinfo {author} {\bibfnamefont {Matthew}\ \bibnamefont
  {Giesler}}, \bibinfo {author} {\bibfnamefont {Daniel~A.}\ \bibnamefont
  {Hemberger}}, \bibinfo {author} {\bibfnamefont {Geoffrey}\ \bibnamefont
  {Lovelace}}, \bibinfo {author} {\bibfnamefont {Kevin}\ \bibnamefont {Kuper}},
  \bibinfo {author} {\bibfnamefont {Michael}\ \bibnamefont {Boyle}}, \bibinfo
  {author} {\bibfnamefont {B.}~\bibnamefont {Szilágyi}}, \ and\ \bibinfo
  {author} {\bibfnamefont {Lawrence~E.}\ \bibnamefont {Kidder}},\ }\bibfield
  {title} {\enquote {\bibinfo {title} {{Improved methods for simulating nearly
  extremal binary black holes}},}\ }\href {\doibase
  10.1088/0264-9381/32/10/105009} {\bibfield  {journal} {\bibinfo  {journal}
  {Class. Quant. Grav.}\ }\textbf {\bibinfo {volume} {32}},\ \bibinfo {pages}
  {105009} (\bibinfo {year} {2015})},\ \Eprint {http://arxiv.org/abs/1412.1803}
  {arXiv:1412.1803 [gr-qc]} \BibitemShut {NoStop}%
\bibitem [{\citenamefont {{LIGO Scientific
  Collaboration}}(2018)}]{aLIGODesignNoiseCurve}%
  \BibitemOpen
  \bibfield  {author} {\bibinfo {author} {\bibnamefont {{LIGO Scientific
  Collaboration}}},\ }\href@noop {} {\emph {\bibinfo {title} {Updated Advanced
  LIGO sensitivity design curve}}},\ \bibinfo {type} {Tech. Rep.}\ (\bibinfo
  {year} {2018})\ \bibinfo {note}
  {\url{https://dcc.ligo.org/LIGO-T1800044/public}}\BibitemShut {NoStop}%
\bibitem [{\citenamefont {{SXS Collaboration}}()}]{SXSCatalog}%
  \BibitemOpen
  \bibfield  {author} {\bibinfo {author} {\bibnamefont {{SXS Collaboration}}},\
  }\href@noop {} {\enquote {\bibinfo {title} {The {SXS} collaboration catalog
  of gravitational waveforms},}\ }\bibinfo {note}
  {\url{http://www.black-holes.org/waveforms}}\BibitemShut {NoStop}%
\bibitem [{\citenamefont {Chatziioannou}\ \emph {et~al.}(2018)\citenamefont
  {Chatziioannou}, \citenamefont {Lovelace}, \citenamefont {Boyle},
  \citenamefont {Giesler}, \citenamefont {Hemberger}, \citenamefont {Katebi},
  \citenamefont {Kidder}, \citenamefont {Pfeiffer}, \citenamefont {Scheel},\
  and\ \citenamefont {Szil\'agyi}}]{Chatziioannou:2018wqx}%
  \BibitemOpen
  \bibfield  {author} {\bibinfo {author} {\bibfnamefont {Katerina}\
  \bibnamefont {Chatziioannou}}, \bibinfo {author} {\bibfnamefont {Geoffrey}\
  \bibnamefont {Lovelace}}, \bibinfo {author} {\bibfnamefont {Michael}\
  \bibnamefont {Boyle}}, \bibinfo {author} {\bibfnamefont {Matthew}\
  \bibnamefont {Giesler}}, \bibinfo {author} {\bibfnamefont {Daniel~A.}\
  \bibnamefont {Hemberger}}, \bibinfo {author} {\bibfnamefont {Reza}\
  \bibnamefont {Katebi}}, \bibinfo {author} {\bibfnamefont {Lawrence~E.}\
  \bibnamefont {Kidder}}, \bibinfo {author} {\bibfnamefont {Harald~P.}\
  \bibnamefont {Pfeiffer}}, \bibinfo {author} {\bibfnamefont {Mark~A.}\
  \bibnamefont {Scheel}}, \ and\ \bibinfo {author} {\bibfnamefont {B\'ela}\
  \bibnamefont {Szil\'agyi}},\ }\bibfield  {title} {\enquote {\bibinfo {title}
  {{Measuring the properties of nearly extremal black holes with gravitational
  waves}},}\ }\href {\doibase 10.1103/PhysRevD.98.044028} {\bibfield  {journal}
  {\bibinfo  {journal} {Phys. Rev. D}\ }\textbf {\bibinfo {volume} {98}},\
  \bibinfo {pages} {044028} (\bibinfo {year} {2018})},\ \Eprint
  {http://arxiv.org/abs/1804.03704} {arXiv:1804.03704 [gr-qc]} \BibitemShut
  {NoStop}%
\end{thebibliography}%

\clearpage
\section*{\large Supplemental materials}
\label{supp_mat}
\renewcommand{\Cornerfignum}{\ref{fig:corner_GW200129}\xspace}
\setcounter{equation}{0}
\setcounter{figure}{0}
\setcounter{table}{0}
\renewcommand{\theequation}{S\arabic{equation}}
\renewcommand{\thefigure}{S\arabic{figure}}
\renewcommand{\thetable}{S\arabic{table}}

\section{High spin NR injections}

\begin{figure*}[thb]
\includegraphics[width=0.49\textwidth]{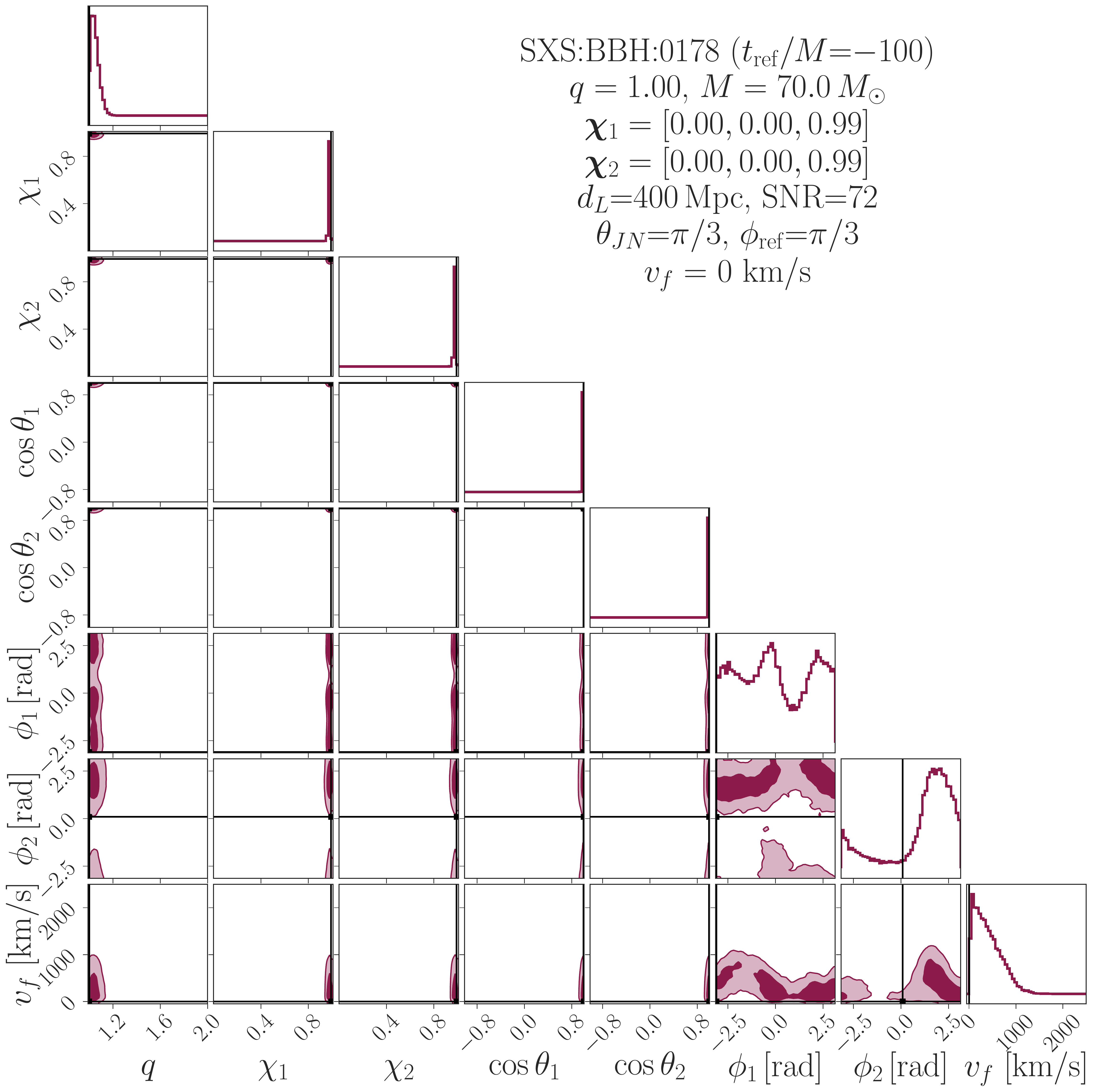}
\includegraphics[width=0.49\textwidth]{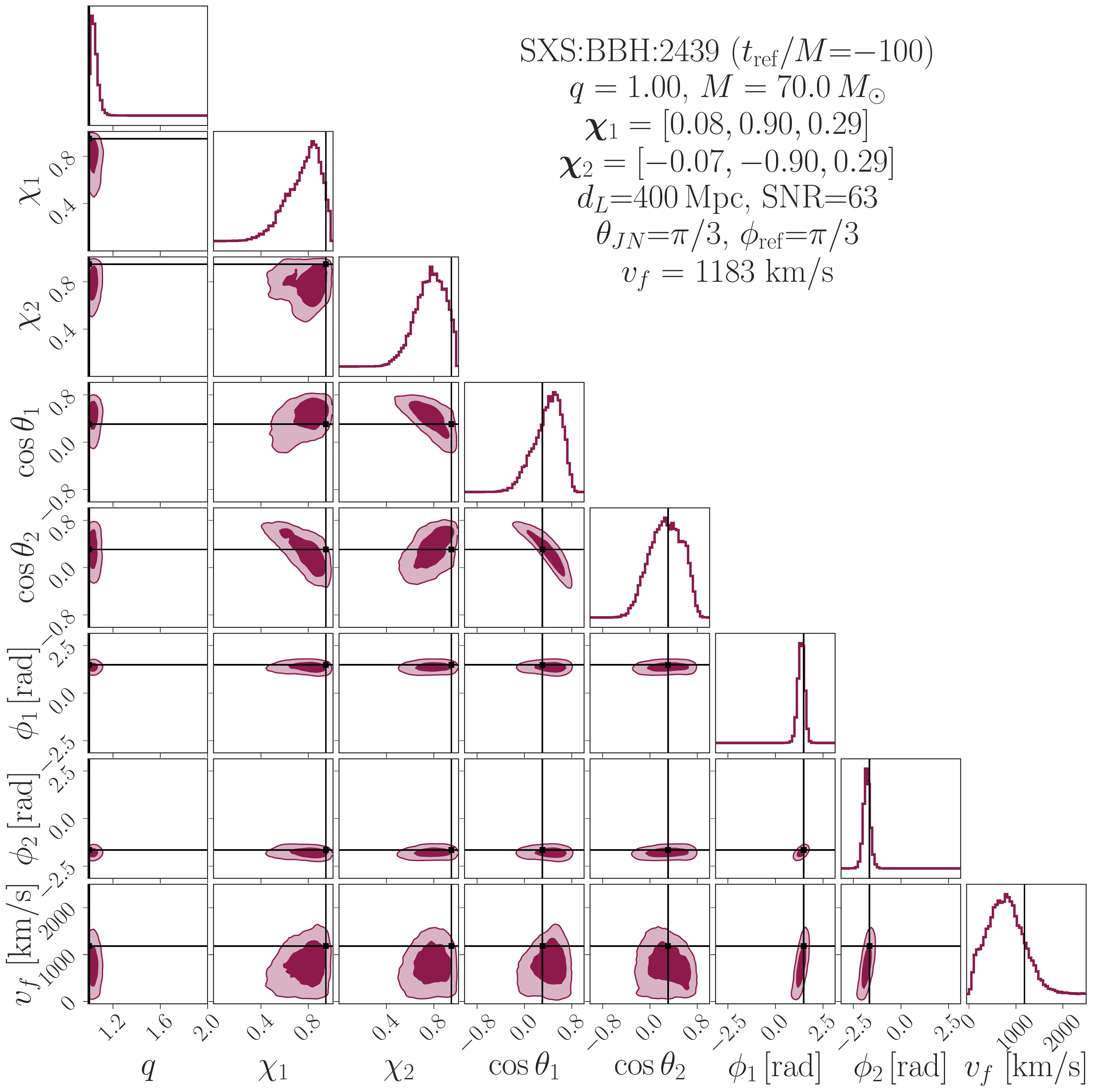}
\caption{
Spin extrapolation tests for the \NRSur and \NRSurRemnant models. Shown are the
recovered kick magnitude, mass ratio and spin posteriors (measured at
$\trefmHundredM$) for NR injections. The dark (light) regions represent the
50\% (90\%) credible bounds on joint 2D posteriors, while the diagonal plots
show 1D marginalized posteriors.  The injected binary parameters are shown as
black lines, and are also given in the inset text. The left panel corresponds
to an equal mass binary with nearly extremal ($\chi_{1,2} \sim 0.99$) but
aligned spins. The right panel corresponds to an equal mass binary in a
superkick configuration, with large spins ($\chi_{1,2} \sim 0.95$). In both
cases, the injected values for the mass ratio, spins, and the kick are well
recovered and lie within the 90\% credible region of the posterior. As the
in-plane spin is zero for the left panel, $\phi_1$ and $\phi_2$ are not
meaningful parameters and the offset from the true value is not of concern.
}
\label{fig:high_spin_inj}
\end{figure*}

As GW200129 has a preference for a large $\chi_1$ (see Fig.~\Cornerfignum), it
is important to check the validity of the surrogate models \NRSur and
\NRSurRemnant in this regime. These models were trained on NR simulations with
$\chi_{1,2} \leq 0.8$, but allow extrapolations to $\chi_{1,2} =
1$~\cite{Varma:2019csw}. Unfortunately, because NR simulations become expensive
for large spins, very few simulations exist beyond $\chi_{1,2} =
0.8$~\cite{Boyle:2019kee}. Therefore, the surrogate models have only been
tested against a handful of NR simulations outside their training
region~\cite{Varma:2019csw, Beth:2021_inprep, Hannam:2021pit}. While an
exhaustive exploration is still prohibitively expensive, we conduct two
additional tests using high spin NR simulations from
Refs.~\cite{Scheel:2014ina, Ma:2021znq}. We inject these NR waveforms (in
zero-noise) into a simulated LIGO-Virgo network operating at design
sensitivity~\cite{aLIGODesignNoiseCurve}. Then, we follow the same procedure as
described in the main text, and use the \NRSur and \NRSurRemnant models to
recover the binary parameters and the kick.

We consider two NR waveforms, SXS:BBH:0178~\cite{Scheel:2014ina} and
SXS:BBH:2439~\cite{Ma:2021znq}, which are publicly available through the SXS
Catalog~\cite{SXSCatalog}. SXS:BBH:0178 corresponds to an equal mass binary
with equal spins that are nearly extremal ($\chi_{1,2} \sim 0.99$), and aligned
along the orbital angular momentum. Because of the symmetries (equal masses and
spins), the kick for this binary is zero. SXS:BBH:2439 also corresponds to a
binary with equal masses and large spins ($\chi_{1,2} \sim 0.95$), but with
most of the spin in the orbital plane. In this case, while the spin magnitudes
are equal, the directions are not. In fact, the spins were chosen to lie in the
``superkick'' configuration~\cite{Ma:2021znq}, with the in-plane spin
components being anti-parallel to each other; this configuration leads to large
kicks~\cite{Campanelli:2007cga, Gonzalez:2007hi, Lousto:2011kp}. For both
injections, we set $d_L=400$ Mpc and $\theta_{JN}=\phi_{\mathrm{ref}}=\pi/3$,
where $d_L$ is the luminosity distance, $\theta_{JN}$ is the inclination angle
between the total angular momentum $\bm{J}$ and line of sight direction
$\hat{\bm{N}}$, and $\phi_{\mathrm{ref}}$ is the reference orbital
phase~\cite{LIGOScientific:2021djp}. This leads to large signal to noise ratios
(SNRs) of 72 and 63, respectively, for SXS:BBH:0178 and SXS:BBH:2439, and
therefore provides a stringent test for the surrogate models. The rest of the
binary parameters will be shown in figure insets below.

Fig.~\ref{fig:high_spin_inj} shows the recovered mass ratio, spins, and kick
for the two NR injections. In both cases, we find that the injected values are
well recovered and lie within the 90\% credible region of the posterior. We
note, however, that the kick posterior for SXS:BBH:2439, while still capturing
the true value, peaks below the injected value, which is related to the fact
that the spin magnitudes also peak below the injected value. Similar trends
were noted in Ref.~\cite{Chatziioannou:2018wqx}, where it was found that the
prior choice of uniform in spin magnitude and direction (which we also adopt in
this work) may not be suitable for binaries with large spins. Therefore, an
investigation on the impact of prior choices on the spin and kick inferences
for high spin events may be necessary, especially as we approach the high SNRs
considered in Fig.~\ref{fig:high_spin_inj} with future detector improvements.
In summary, while these tests give us more confidence in the constraints placed
on GW200129, we recommend a more exhaustive study to rule out any potential
systematics in the surrogate models due to extrapolation to large spins, which
we leave for future work.

\end{document}